\documentclass[conference]{IEEEtran}

\ifCLASSOPTIONcompsoc

  \usepackage[nocompress]{cite}
\else
  \usepackage{cite}
\fi

\ifCLASSINFOpdf

\else

\fi

\IEEEoverridecommandlockouts
\makeatletter\def\@IEEEpubidpullup{6.5\baselineskip}\makeatother
\IEEEpubid{\parbox{\columnwidth}{
    Network and Distributed System Security (NDSS) Symposium 2025\\
    24–28 February 2025, San Diego, CA, USA\\
    ISBN 979-8-9894372-8-3\\
https://dx.doi.org/10.14722/ndss.2025.242185\\
www.ndss-symposium.org
}
\hspace{\columnsep}\makebox[\columnwidth]{}}

\usepackage{graphicx}
\usepackage{caption}
\usepackage{amsmath}
\usepackage{subcaption}
\usepackage{balance}
\usepackage{tabularx}
\usepackage{booktabs}
\usepackage{microtype}[final]
\usepackage{graphicx} 
\usepackage{amsmath,amsfonts}
\usepackage{algorithmic}
\usepackage{graphicx}
\usepackage{textcomp}
\usepackage{xcolor}
\usepackage{fancyhdr}
\usepackage{hyperref}
\usepackage{algorithm}
\usepackage{algorithmic}
\usepackage{caption}
\usepackage{subcaption}
\usepackage{amsmath,amsfonts}
\usepackage[braket, qm]{qcircuit}
\usepackage{multirow}
\usepackage{algorithm}
\usepackage{algorithmic}
\usepackage{graphicx}
\usepackage{enumitem}
\usepackage{textcomp}
\usepackage[export]{adjustbox}
\usepackage{textcomp}
\usepackage{braket}
\usepackage{amsmath}
\usepackage{xcolor}
\usepackage{graphicx}

\hypersetup{ colorlinks=true, linkcolor=black, urlcolor=black, filecolor=black, linkbordercolor=white, citecolor=black}

\begin{document}

\title{Crosstalk-induced Side Channel Threats in Multi-Tenant NISQ Computers}

\author{\IEEEauthorblockN{Navnil Choudhury\IEEEauthorrefmark{1},
Chaithanya Naik Mude\IEEEauthorrefmark{2},
Sanjay Das\IEEEauthorrefmark{1}, 
Preetham Chandra Tikkireddi\IEEEauthorrefmark{2},
Swamit Tannu\IEEEauthorrefmark{2} \\and 
Kanad Basu\IEEEauthorrefmark{1}}
\IEEEauthorblockA{\IEEEauthorrefmark{1}
University of Texas at Dallas,
Richardson, Texas, \IEEEauthorrefmark{2}University of Wisconsin-Madison, Madison, Wisconsin}}

\maketitle

\IEEEpeerreviewmaketitle

\begin{abstract}

As quantum computing rapidly advances, its near-term applications are becoming increasingly evident. However, the high cost and under-utilization of quantum resources are prompting a shift from single-user to multi-user access models. In a multi-tenant environment, where multiple users share one quantum computer, protecting user confidentiality becomes crucial. The varied uses of quantum computers increase the risk that sensitive data encoded by one user could be compromised by others, rendering the protection of data integrity and confidentiality essential.
In the evolving quantum computing landscape, it is imperative to study these security challenges within the scope of realistic threat model assumptions, wherein an adversarial user can mount practical attacks without relying on any heightened privileges afforded by physical access to a quantum computer or rogue cloud services. 



In this paper, we demonstrate the potential of crosstalk as an attack vector for the first time on a Noisy Intermediate Scale Quantum (NISQ) machine, that an adversarial user can exploit within a multi-tenant quantum computing model.
The proposed side-channel attack is conducted with minimal and realistic adversarial privileges, with the overarching aim of uncovering the quantum algorithm being executed by a victim. Crosstalk signatures are used to estimate the presence of CNOT gates in the victim circuit, and subsequently, this information is encoded and classified by a graph-based learning model to identify the victim quantum algorithm. When evaluated on up to 336 benchmark circuits, our attack framework is found to be able to unveil the victim's quantum algorithm with up to 85.7\% accuracy.

\end{abstract}

\section{Introduction}
\label{sec:intro}

Quantum computing is attracting interest for its ability to perform calculations much faster than traditional computers. This speed is showcased in algorithms like Shor's and Grover's, which significantly enhance computational efficiency~\cite{grover1996fast,Shor}. Moreover, quantum computers promise to enable efficient simulations of quantum systems such as molecules and materials, paving the path toward designing new drugs, batteries, fertilizers, and even more~\cite{troyer}. At its core, quantum computing uses algorithms built with quantum circuits that consist of qubits—the basic building blocks of quantum information. Through quantum gates, the states of these qubits are manipulated, allowing quantum computers to perform complex calculations and transformations that are beyond the computational capabilities of traditional computers.

\begin{figure*}[t]
 \centering
 \begin{subfigure}[b]{0.26\textwidth}
 \includegraphics[width=\textwidth]{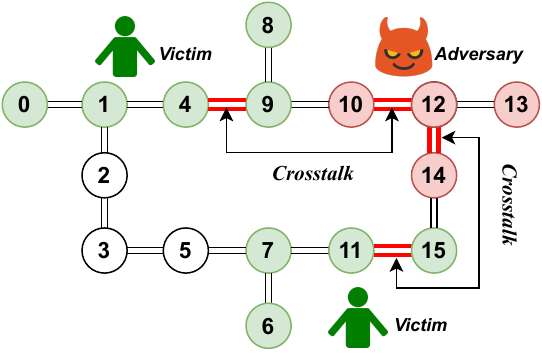}
 \caption{Adversary inducing crosstalk-based side channel. }
 \label{qec}
 \end{subfigure}
 \hfill
 \begin{subfigure}[b]{0.33\textwidth}
 \includegraphics[width=\textwidth]{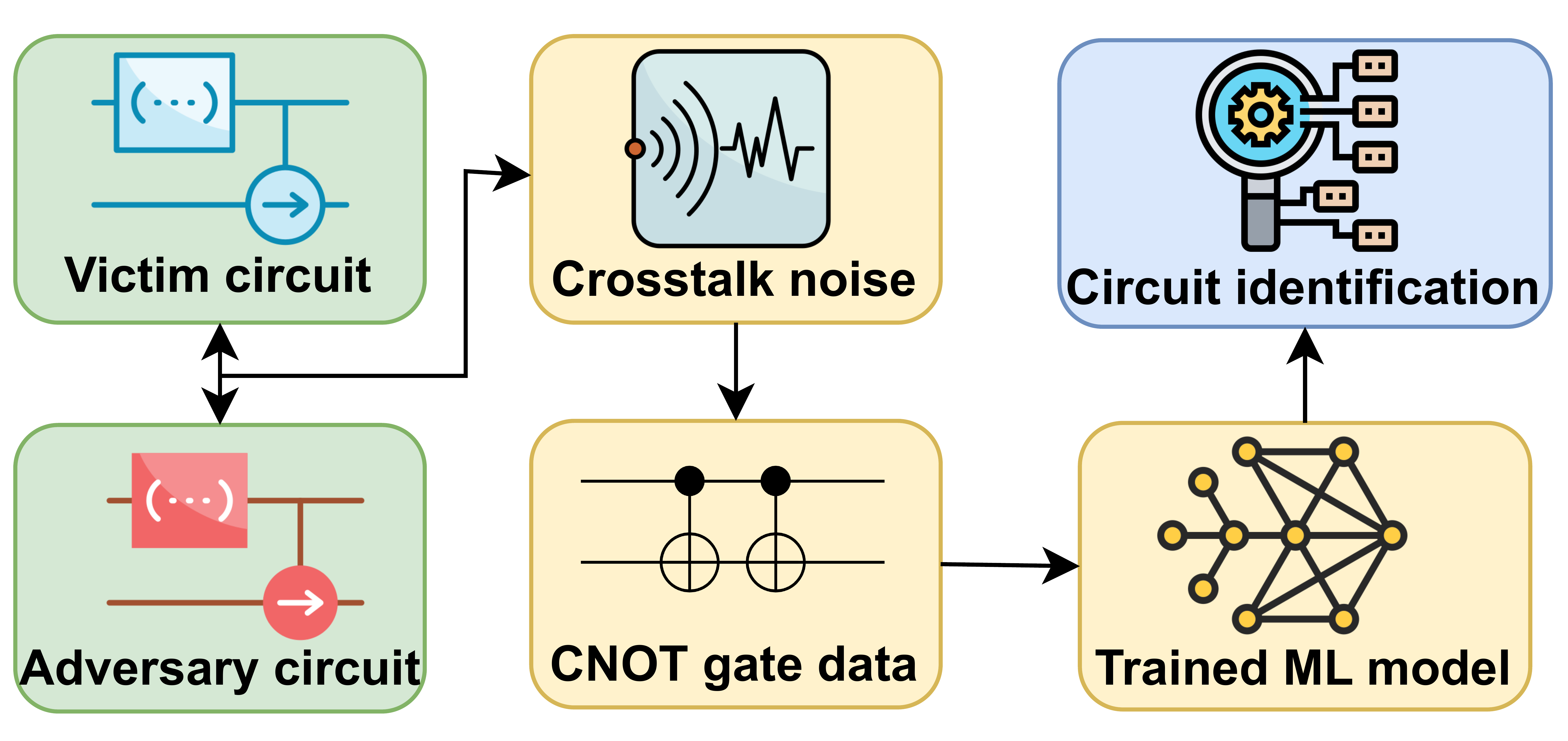}
 \caption{Our proposed attack framework under real-world constraints.}
 \label{quantumcloud}
 \end{subfigure}
 \hfill
 \begin{subfigure}[t]{0.36\textwidth}
 \vspace{-30mm}
\resizebox{\linewidth}{!}{%
\begin{tabular}{c|c|c}
\hline
\textbf{Existing approach} & \textbf{Adversary's objective}                                            & \textbf{Threat model limitations}                                                                              \\ \hline
Saki \textit{et al.}   (arxiv 2021)             & \begin{tabular}[c]{@{}c@{}}Qubit state\\  recovery\end{tabular}           & \begin{tabular}[c]{@{}c@{}}Co-location on quantum chip is \\ not guaranteed.\end{tabular}                \\ \hline
Deshpande \textit{et al.}   (CCS 2022)        & \begin{tabular}[c]{@{}c@{}}Quantum circuit \\ detection\end{tabular}      & \begin{tabular}[c]{@{}c@{}}100 \% qubit utilization is \\ not feasible.\end{tabular}                           \\ \hline
Erata \textit{et al.}      (CCS 2023)        & \begin{tabular}[c]{@{}c@{}}Quantum circuit \\ reconstruction\end{tabular} & \begin{tabular}[c]{@{}c@{}}Physical access required to quantum \\ computers is highly restricted.\end{tabular} \\ \hline
\end{tabular}%
}
\vspace{5mm}
\caption{Existing threat models on QaaS systems and their limited feasibility. }
 \label{threatmodel}
\end{subfigure}
 
 \caption{Figure \ref{qec} demonstrates an adversary using sparse connections in the coupling architecture of IBM Quantum devices to induce crosstalk. Figure \ref{quantumcloud} depicts our proposed attack model, which leverages this crosstalk to identify the victim circuit. Figure \ref{threatmodel} highlights the limitations of existing threat models, emphasizing the need for a practical threat model.}
 \label{fig:rq2_1}
 \vspace{-6mm}
 \end{figure*}



Despite their numerous advantages, quantum computers are expensive. They need precise instrumentation and cryogenic operating environments, rendering it infeasible for most users to obtain dedicated access to a quantum computer. This poses a significant barrier to the widespread accessibility of quantum computers for the foreseeable future. In order to democratize quantum computing and make it accessible, there has been a significant rise in interest and practical deployments of remote cloud-based quantum computing services, commonly known as Quantum as a Service (QaaS). These services aim to provide remote access to quantum computers, enabling users to harness the power of quantum computation without the need for local infrastructure or expertise. 
QaaS supports numerous near-term applications across various domains of scientific research, including but not limited to biochemistry, cryptography, finance, and machine learning~\cite{chemestry,optimization,lu2022survey}.




Considering the diverse range of applications of QaaS, it is important to recognize that users leveraging QaaS may be utilizing quantum algorithms that involve sensitive encoded data on these cloud-based quantum computers. 
Moreover, many QaaS vendors, such as IBM, are rapidly expanding their quantum computing capabilities, now offering access to systems with over 1000 qubits, with doubling qubit counts annually~\cite{Castelvecchi2023}.
Consequently, it becomes essential to prioritize the security of the computational data uploaded by users.

Among the various security challenges, the threat of side-channel leakage risks in quantum computers is especially heightened due to a fundamental issue: \textbf{quantum programs or circuits produce a probability distribution as the output, and each execution run provides only a single sample}. Thus, all quantum circuits must be repeatedly executed to obtain the final output.  Therefore, the co-located adversary can snoop not just one, but a large number of quantum circuit runs. 
To this end, Figure \ref{fig:rq2_1} provides an overview of the operational framework of our proposed side-channel information leakage attack on a QaaS within a multi-tenant environment. 

First, Figure \ref{qec} depicts a potential security violation in a 16-qubit NISQ machine developed. This attack leverages crosstalk in such NISQ machines, the details of which are covered in Section \ref{sec:xtalk}.
Subsequently, Figure \ref{quantumcloud} depicts an overview of our proposed attack, which entails determining the position and location of two-qubit operations such as CNOT gates, and using this information to subsequently identify the victim quantum circuit. The details of the relation between crosstalk and CNOT gate information are provided in Section \ref{sec:xtalk}, and the construction of the graph-based classifier used in this endeavor is highlighted in Section \ref{sec:gnn}. Finally, the limitations of prevalent threat models based on existing NISQ machines (specifically on IBM superconducting devices) are depicted in Table \ref{threatmodel}, and are discussed further in Section \ref{sec:related}.
Despite the significant progress, the prevalent QaaS systems often face under-utilization, resulting in long wait times for users. This is primarily due to the single-tenant model used, where a quantum computer is dedicated to just one user at a time, leading to inefficiencies in system utilization.
To address these limitations and manage the escalating demand for quantum computing resources, cloud-based quantum computers are shifting towards the adoption of a multi-tenant model of computing~\cite{das2019case,niu2023enabling}. 
This paradigm shift improves throughput by enabling the concurrent execution of diverse programs, allocating distinct sets of qubits in a quantum computer to multiple users. Consequently, the utilization of available quantum resources could be maximized, resulting in enhanced system efficiency.
The transition towards a multi-tenant model increases the system's capacity to accommodate a larger user base by allowing multiple users to access quantum computers simultaneously. 
However, a multi-tenant framework incurs several operational challenges~\cite{das2019case}. These challenges primarily arise due to the presence of noise, which has a pronounced effect on the current generation of quantum systems, giving rise to the \textit{``Noisy Intermediate Scale Quantum"}, \textbf{\textit{(NISQ)}} era.

The broad accessibility of multi-tenancy-enabled QaaS can lead to potential adversarial activities targeting fellow users or the underlying infrastructure. Malicious users can initiate attacks, leveraging the prevalent crosstalk inherent in NISQ-era computers.
Consequently, there exists the inherent pitfall of unauthorized information leakage, whereby a malicious user may attempt to illicitly access and extract sensitive data pertaining to other users sharing the same infrastructure. Due to the growing applications of quantum computing, they are used for solving numerous problems, including drug discovery and financial portfolio optimization, where sensitive and proprietary information is encoded onto quantum circuits, making security breaches a significant concern. As an example of such a scenario, consider a pharmaceutical company simulating unique molecular structures for drug discovery problems, where proprietary quantum circuits are executed. In such a scenario, unauthorized access to circuit information can lead to theft of intellectual property, having significant consequences.

Specifically, identifying a user's quantum circuit poses a significant threat in quantum computing, setting it apart from risks encountered in classical computing paradigms. For example, identifying an Advanced Encryption Standard (AES) algorithm by side-channel analysis on classical hardware has limited impact unless the AES key is disclosed. Therefore, sensitive information remains safeguarded even if circuit details are exposed. However, during quantum circuit design, sensitive data is encoded into qubits and the quantum circuit, which can include confidential details. Identifying the quantum circuit allows for the extraction of this sensitive information. For example, in quantum chemistry applications, such as the Hartree-Fock method for finding approximate solutions to the Schrödinger equation, CNOT gates are used to prepare the quantum state of a molecule~\cite{google2020hartree}. The quantum state encodes the electronic configuration of the molecule. If an adversary can determine the circuit used, deducing the type of molecule becomes possible, thereby revealing highly sensitive information and trade secrets.

Existing research has explored the extraction of sensitive information from a target (victim) circuit within a multi-tenancy enabled NISQ quantum computing system, attributed to an adversarial entity and characterized as a side-channel attack~\cite{erata2024quantum, 10.1145/3370748.3406570}.
\textbf{However, the efficacy of such attacks is contingent upon stringent assumptions, including unrestricted physical access to the quantum computing infrastructure, the adoption of simplified parametric circuits for the targeted circuit, or highly escalated operational privileges for the adversary}~\cite{10.1145/3576915.3623118,9951250, 10133711, erata2024quantum}. 
From a quantum cloud services perspective, existing attack frameworks and threat models are impractical. Therefore, we need a deeper understanding of vulnerabilities and their possible exploitation within multi-tenant quantum systems. To that end, we present a comprehensive and reproducible threat model applicable under real-world constraints, underscored in Section \ref{sec:threat}.

To launch a practical attack, we could leverage information leakage induced by operational crosstalk~\cite{PRXQuantum.3.020301}. 
On quantum computers, crosstalk refers to the undesired interaction or interference between qubits in a quantum computing system. It occurs when the state or operation of one qubit influences the state or operation of another qubit, leading to errors or inaccuracies in quantum computations. This is depicted in Figure \ref{qec}, where the operations in qubit pairs \{\textit{4,9}\} and \{\textit{11,15}\} are affected by crosstalk emanating from qubit pairs \{\textit{10,12}\} and \{\textit{12,14}\}, respectively. 
The effects of quantum crosstalk can manifest in several ways, resulting in unwanted entanglement or correlations between qubits. This can corrupt quantum information, thereby affecting circuit fidelity. The potential for crosstalk as an attack vector has been identified through readout crosstalk, which aims to steal the output data produced by quantum algorithms~\cite {maurya2024understanding}. In this paper, our proposed approach, for the first time, shows that operational crosstalk can lead to information leakage, thereby {\em enabling circuit stealing attacks}, as explained in Section \ref{sec:xtalk}.

In this paper, for the first time, we present an end-to-end circuit-stealing attack by exploiting crosstalk between qubits. 
The proposed attack operates under minimal adversary privileges, highlighting the significance of resource sharing in compromising system security, as shown in Figure~\ref{qec}. The proposed attack targets the connectivity of a specific user's circuit, referred to as the victim. Subsequently, a graph convolutional network (GCN)-based classification model is used to identify the victim's circuit, thereby compromising the victim's confidentiality.
The attack is designed to extract enough information from a victim circuit to be able to reconstruct it, demonstrating the potential threat of concurrent execution in a multi-tenant environment. Since the victim circuit might contain sensitive information, such an attack can have disastrous consequences since the victim circuit itself comprises sensitive encoded information in the quantum domain.

The key contributions of the paper are:
\begin{itemize}[noitemsep,topsep=0pt]
    \item To the best of our knowledge, this is one of the first works that extensively explores the potential vulnerabilities in a multi-tenant quantum computer using a crosstalk-based attack vector within the scope of a realistic threat model.

    \item We introduce a novel end-to-end attack that capitalizes on the operational crosstalk in quantum circuits, revealing highly sensitive information about the circuit's structure.  Specifically, we demonstrate it is possible to recover the number of CNOT gates utilized in a victim circuit. 

    \item We illustrate the relationship between the connectivity created by CNOT gates in a quantum circuit, and the construction of quantum algorithms, further demonstrating the capability to identify a quantum algorithm based on its underlying CNOT gate connections.
    
    \item Finally, we extract relevant factors pertaining to CNOT gate connectivity in quantum circuits and encode them as features. Following this, we utilize a GCN-based classification model for successfully identifying the victim circuit, encoding the extracted information as features. The model attains an accuracy of 85.7\% when evaluated on 336 benchmark circuits.
\end{itemize}

\section{Background and motivation} \label{sec:background}

This section provides an overview of the basic quantum computing concepts, in addition to side-channel vulnerabilities in quantum computers.   


\subsection{Quantum Circuits, and Noise} 

\paragraph{\textbf{Qubits}}
The fundamental building block of a quantum computer is the qubit, analogous to a classical bit in classical computing. A qubit can be mathematically represented as a two-dimensional column vector with a unit norm. Qubit state can be expressed as a linear combination of basis states $\ket{0}$ and $\ket{1}$, which are defined as:
$$
\ket{0} = \begin{bmatrix}1\\0\end{bmatrix}\hspace{0.5cm}
\ket{1} = \begin{bmatrix}0\\1\end{bmatrix} \hspace{0.5cm}
\ket{\psi} = \alpha \ket{0} + \beta \ket{1} = \begin{bmatrix}\alpha\\\beta\end{bmatrix} $$

When a qubit device that holds quantum state $\ket{\psi}$ is measured, it yields $\ket{0}$ with probability $\alpha^{2}$ or $\ket{1}$  with probability $\beta^{2}$. Qubit measurement is probabilistic and destructive, upon measuring a qubit, a binary outcome is produced, collapsing a quantum superposition to $\ket{0}$ or $\ket{1}$. Therefore, to infer the output of a quantum circuit, we need to repeat the execution of the quantum circuit and the qubit measurement.



\paragraph{\textbf{Quantum Gates}} 
Quantum gates are used to manipulate the state of quantum bits. Quantum gates operate on single or multiple qubits and can be expressed mathematically as a matrix with dimensions of $2^n\times 2^n$, where $n$ corresponds to the number of qubits they operate upon. Diverse quantum computing interfaces provide a variety of quantum gates, including but not limited to the X-Gate, H-Gate, CNOT Gate, \textit{etc}. Note that on most quantum hardware platforms, only the single and two-qubit gates are supported, and larger gates are decomposed into a sequence of single and two-qubit gates. 

\par
$X =  
\begin{bmatrix}
  0 & 1 \\
  1 & 0
\end{bmatrix}$
$H = 
\begin{bmatrix}
    \frac{1}{\sqrt{2}} & \frac{1}{\sqrt{2}} \\
    \frac{1}{\sqrt{2}} & \frac{-1}{\sqrt{2}}
\end{bmatrix}$
$CNOT = 
\begin{bmatrix}
    1 & 0 & 0 & 0 \\
    0 & 1 & 0 & 0 \\
    0 & 0 & 0 & 1 \\
    0 & 0 & 1 & 0
\end{bmatrix}$
\smallskip

CNOT gates, which create entanglement between physical qubits, are particularly important in quantum circuits. Understanding the sequence of these {\em CNOT gates can reveal a lot about the structure of a quantum circuit. Since CNOT gates significantly influence the overall behavior and functionality of the circuit, identifying them can potentially lead to the identification of the entire circuit.}

\paragraph{\textbf{Quantum Circuits}} A quantum circuit is a sequence of quantum gates applied to the collection of qubits designed to attain specific logical outcomes. 
Intuitively, a quantum circuit operating on \textit{N} qubits gives us the ability to manipulate the quantum state with $2^N$ possible outcomes. For example, a circuit operating on four qubits can create a quantum state that is a superposition of sixteen states (``0000'' to ``1111''). Depending upon the circuit, these four-bit states appear with a certain probability. However, when we measure four qubits at the end of the circuit, the result is a 4-bit binary string. Thus, we must execute the circuit multiple times to infer the quantum state that encodes the probability distribution (\textit{i.e.}, the output of the circuit)\footnote{Most NISQ applications require thousands of circuit runs~\cite{tannu2019not}. }. Essentially, a sequence of quantum gate operations creates a probability distribution, and the act of measurement is akin to sampling from this distribution. {\em Repeated executions enhance the effectiveness of potential side channels by giving adversaries multiple opportunities to observe the circuit execution.}   

A quantum algorithm consists of a series of instructions designed to solve a specific problem, represented by a quantum circuit. The circuit applies quantum gates to qubits, leveraging phenomena like superposition and entanglement. Each quantum circuit corresponds to a particular quantum algorithm, defining its computation process. However, a single quantum algorithm can be implemented through multiple quantum circuits, each with different gate arrangements or structures. By analyzing the types and sequence of quantum gates in a circuit, one can gain insight into which specific algorithm is being executed. 

\paragraph{\textbf{Noisy Quantum Hardware}} Qubits are fickle; even a small perturbation in the environment can change the state of a qubit. Errors in quantum computers can be classified into three categories: coherence errors, operational errors, and crosstalk errors. Our threat model specifically deals with the security implications associated with crosstalk errors. Performing a gate operation on a qubit can affect the state of its neighboring qubits, a phenomenon known as crosstalk error~\cite{sarovar2020detecting}. 
Most qubit technologies suffer from crosstalk errors~\cite{tripathi2021suppression,murali2020software,das2021adapt}. {\em Crosstalk errors are not confined to the qubits that the user is operating and can affect the remote and even non-operational qubits, leading to information leakage.}







\subsection{Quantum Cloud and  Multi-tenancy} \label{subsec:multcloud}
Quantum computers are expensive due to their need for precise control and operation in extremely cold temperatures. To make them accessible to more users, a cloud-based service known as Quantum-as-a-Service (QaaS) is used. This service simplifies the complex process of controlling qubits for users.

Companies like Amazon, IBM, and Microsoft offer QaaS\cite{qiskit, braket}. Users send their quantum circuits and the desired number of iterations to these providers. The cloud providers then run these circuits on quantum hardware and return the results. Users pay based on their usage time on the cloud. Quantum circuits need to be run multiple times (thousands of shots) to get accurate results, and this process is slow. A single run of a quantum circuit can take milliseconds, and repeating it thousands of times can extend to several seconds.

As more people use QaaS, the wait times increase because there are relatively few quantum computers. Currently, each user gets exclusive access to all qubits during their session, even if they don't use all of them. This leads to inefficient use of qubits. To improve efficiency, there's a growing trend towards a multi-tenant model in quantum computing. Unlike the current single-tenant system, where all qubits are allocated to a single user for the duration of their program execution and the qubits are shared fairly in time, the multi-tenant model allows several users to use disjoint sets of qubits on the quantum computer simultaneously. This approach will significantly improve the utilization of scarce quantum resources.

\subsection{Potential Side Channel Targets on Quantum Computer}

\begin{figure}[t]
\centering
 \includegraphics[width=0.9\linewidth]{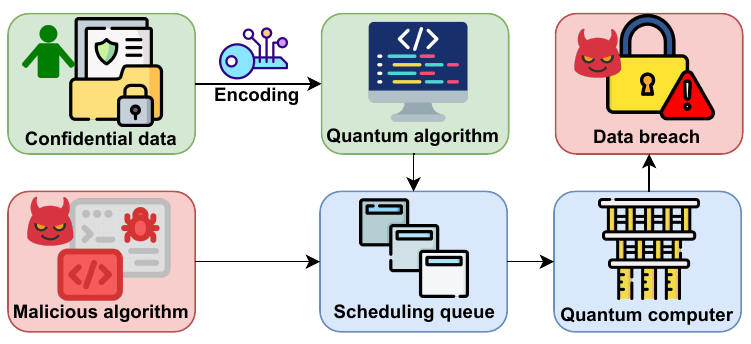}
 \caption{Operational attack for confidential information extraction in prevalent QaaS environment. }
 \label{fig:endtoend}
 \vspace{-5mm}
 \end{figure}

Researchers in both industry and academia use quantum cloud platforms to develop applications and explore the potential of quantum computers for solving real-world problems. The quantum circuits they create are valuable intellectual property and can contain sensitive information. If these circuits are exposed, it could lead to a breach of this intellectual property. Figure \ref{fig:endtoend} demonstrates the operational flow of such an attack causing a breach of security.

Particularly important in quantum circuits are CNOT gates, which create connections between physical qubits. Understanding the sequence of these CNOT gates can reveal a lot about the structure of a quantum circuit. Since CNOT gates significantly influence the overall behavior and functionality of the circuit, identifying them can potentially lead to the identification of the entire circuit. The patterns formed by the connections between qubits, established by CNOT gates, provide crucial structural information that can be used to understand the circuit more deeply, gleaning sensitive information in the quantum circuit.

This paper focuses on exploring the risk of crosstalk in multi-tenant quantum computing environments. Crosstalk occurs when multiple quantum circuits interfere, which can happen when different users share the same quantum computing resources. To this end, we proceed to develop a comprehensive threat model (outlined in Section \ref{sec:threat}) to assess the potential security vulnerabilities associated with crosstalk, especially those related to CNOT gates and idle qubits within a shared quantum computing environment.

\section{Related Work} \label{sec:related}

Crosstalk errors have been a prevailing concern in the domain of quantum computing ever since its inception, prompting extensive investigation in numerous scholarly works~\cite{10.1145/3373376.3378477, 10.1145/3370748.3406570}. The research in this area has encompassed a thorough examination of the challenges arising from crosstalk, as well as the development and evaluation of mitigation strategies to address these issues. 
However, these existing works primarily address crosstalk for a single-tenant model of computing. As such, the mitigation strategies do not extend to a multi-tenant model of computing.

Potential threat models exploiting existing errors prevalent in quantum computers have been studied recently. QubitSensing has explored the possibility of recovery of the state of a qubit following the completion of its operation in a quantum computer\cite{saki2021qubit}. However, the authors assume the co-location of the adversarial qubit with the victim qubit in the quantum computer. This is a strong assumption, making the threat model itself unviable. In addition, this strategy is only applicable for partial information extraction from a victim quantum circuit.

The design of a quantum computer antivirus has been explored in another paper\cite{10133711}. This paper proposes the identification of malicious circuits from a database of circuit sequences that have proven to be malicious. This framework considers an improbable threat model with 100\% utilization of qubits present in a quantum computer~\cite{ash2019qure}. It also suffers from the drawback of not being able to successfully mitigate any sequence it cannot identify, which renders this defense unfeasible for the foreseeable future.

Following this, a power side-channel attack on a quantum computer has been proposed in a recent study\cite{erata2024quantum}. This study focuses on utilizing pulse signatures of quantum gates, which are directly correlated with the power consumption of a circuit. This enables the identification of the type of operations being executed by a circuit, or in some cases, it might lead to potentially identifying the entire circuit. This attack however requires a malicious insider with physical access to the quantum computer, rendering it as a very weak threat model.

\section{Threat Model} \label{sec:threat}

Our threat model makes minimal assumptions about adversarial privileges.\textit{ Unlike prior quantum side-channel attacks, our threat model does not require a rogue cloud or an insider adversary in the cloud services}~\cite{erata2024quantum,10.1145/3373376.3378477}.
In order to analyze the problem and develop a realistic attack leveraging crosstalk-based quantum side channels, we specify the objective of the attacker and outline the assumptions of the attackers' access.

\noindent{\underline{\textit{Assumptions on Attacker's Objective}}:} 
We assume that the attacker's primary objective is to extract critical information regarding the number of CNOT gates employed in the victim circuit, including their precise timing of execution. Consequently, the attacker intends to utilize this information to gain complete knowledge of the victim circuit, including its design, gate operations, and qubit interactions.
The attacker aims to obtain at least partial information about the victim circuit, precisely an estimate of qubit connectivity, which can be utilized to reconstruct the circuit partially or approximately, thereby compromising the victim's confidentiality.


\noindent{\underline{\textit{Assumptions of Attacker's access}}:}
We assume a scenario where both the victim and the adversary can execute their respective quantum programs on the same quantum computer, but on disjoint sets of qubits. \textbf{It is important to note that our threat model has no requirement for co-location between the victim and the adversary's quantum circuit.}
This strict assumption enables us to devise a viable attack strategy on multi-tenant NISQ machines under real-world constraints, with limited access within the QaaS cloud environment. 
The adversary, like any other user in a multi-tenant environment, is assumed to have the capability to manipulate the state and operations of the qubits within the scope of their program. 



\noindent{\underline{\textit{Assumptions of Attacker's Advantages}}:}
The attacker is provided with three minimal and realistic advantages. (1) First, they are assumed to be able to execute their program using qubits in the same quantum computer as the victim circuit. This is a justified assumption in a multi-tenant quantum computer.
\textit{It is worth noting that simply removing or relocating the adversary's quantum circuit from the designated quantum computer by the cloud provider presents several challenges outlined as follows. Identifying the adversarial program to remove is challenging due to the vast number of potential quantum circuits the adversary could construct by varying the delay times between circuit operations for the attack. Furthermore, in the unlikely event that the adversarial quantum circuit is relocated to a different set of qubits, it retains the capability to extract information regardless.
This assumption enables potential interactions and entanglements between the attacker's qubits and the victim circuit's qubits (as demonstrated in Section \ref{sec:xtalk})}.
(2) Secondly, the attacker is assumed to be aware of the start timing or execution phases of the victim circuit. This allows the attacker to concurrently execute their program along with the victim's.
\textbf{This is reasonable since any quantum circuit is executed thousands of times on the QaaS cloud} (as explained in Section \ref{subsec:multcloud}).  
(3) Finally, we assume that the number and types of circuits the victim can execute on their qubits are limited.\textbf{ This is a justified assumption since the aim of the adversary is to steal sensitive information, and practically useful quantum algorithms are limited today.} Furthermore, hardware restrictions imposed on prevalent QaaS systems serve to further limit deployable quantum algorithms, thereby justifying our assumption~\cite{PRXQuantum.2.040335}.

By explicitly stating these assumptions, we establish a threat model that includes the attacker's capabilities, objectives, and limited knowledge in the context of compromising the confidentiality of the victim circuit. 
This includes the capability of the adversary to exploit quantum crosstalk to gain sensitive information encoded in the victim circuit. This endeavor is discussed in detail in Section \ref{sec:gnn}.

\section{Crosstalk-based Side Channels on Quantum Computers} \label{sec:xtalk}

This section will discuss how crosstalk manifests on quantum hardware and how it can be exploited to enable side channels on multi-tenant quantum computers. First, we will explain the crosstalk on superconducting quantum hardware and then explain how we use IBM's quantum hardware to prototype the side channel.

\subsection{Crosstalk on Quantum Hardware} \label{subsec:cross}
Crosstalk errors on quantum computers occur due to unwanted interactions between various components of the quantum hardware, such as the qubits themselves, the electronics that control them, and even parts of the cooling system~\cite{sarovar2020detecting}. Many environmental factors can influence these interactions; some are fundamentally unavoidable. There are two main types of crosstalk errors: operational and idling. Operational crosstalk occurs when multiple operations are carried out at the same time, leading to interference and incorrect results~\cite{GST_crosstalk,murali2020software,das2021jigsaw}. Idling crosstalk occurs when an active gate operation inadvertently impacts a qubit that is not currently being used, altering its state~\cite{das2021adapt}.    


\begin{figure}[t]
    \centering
    \includegraphics[width=0.9\linewidth]{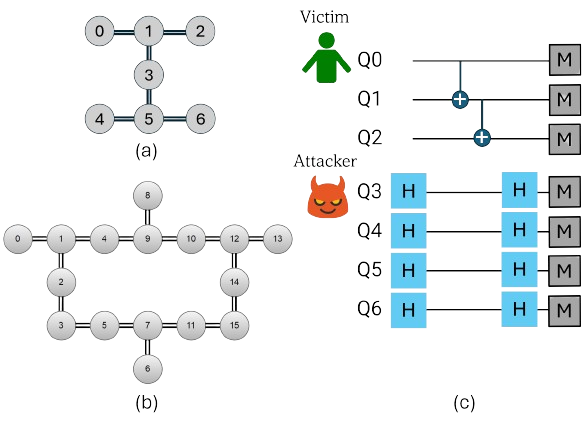}
    \vspace{-0.1in}\caption{Qubit connectivity map for (a)\texttt{IBM Lagos} machine (b)\texttt{IBM Guadalupe} machine (c) Representative crosstalk detection circuit.}
    \label{fig:crosstalk_setup}
    \vspace{-0.2in}
\end{figure}

\begin{figure}[t]
    \centering
    \includegraphics[width=\linewidth]{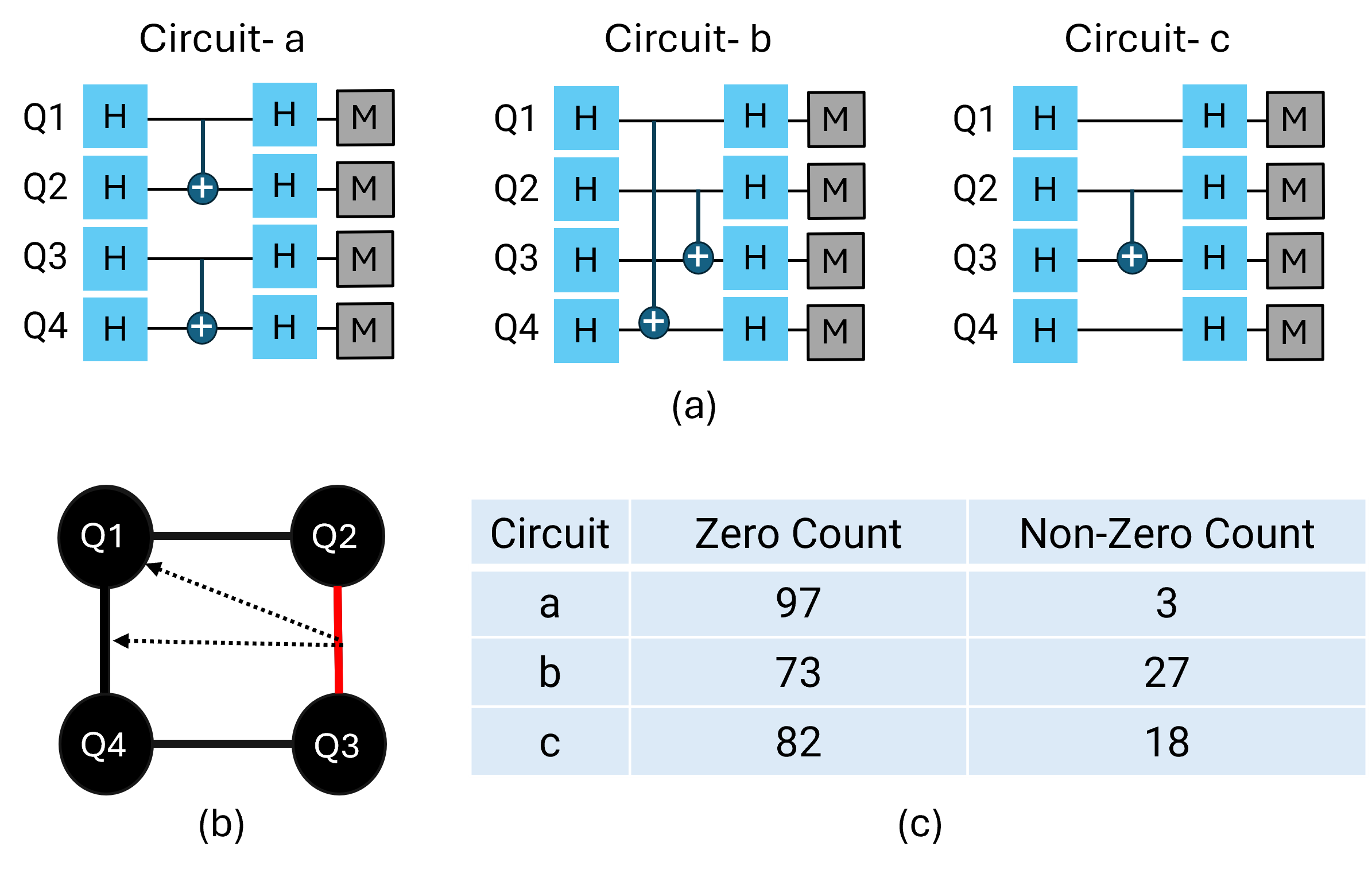}
    \vspace{-0.2in}
    \caption{(a) Crosstalk Detection Circuits (b) Qubit Connectivity Map (c) Zero counts that quantify the level of crosstalk.}
    \label{fig:crosstalk_example}
    \vspace{-0.1in}
\end{figure}

Figure~\ref{fig:crosstalk_example} illustrates how crosstalk impacts quantum programs (or circuits). It displays three quantum circuits, each using four qubits labeled Q1, Q2, Q3, and Q4. These circuits are run on a quantum computer where the four qubits are interconnected, as depicted in Figure~\ref{fig:crosstalk_example} (b). In this figure, the qubits are shown as nodes and their connections, crucial for two-qubit operations, as edges. A two-qubit operation between Q3 and Q2, highlighted in red, causes crosstalk that adversely affects Q1. This interference can increase the error rate in a simultaneous operation involving Q1 and Q4. This crosstalk typically results from `\textit{frequency crowding}', where multiple qubits have similar resonance frequencies \cite{hertzberg2021laser}.

To estimate crosstalk, one typically employs detailed quantum metrology protocols. However, for basic detection of crosstalk, simpler characterization circuits, as illustrated in Figure~\ref{fig:crosstalk_example}(a), can be used. Without crosstalk, these circuits should ideally output a `0000' state with high probability. In contrast, the presence of operational crosstalk leads to non-zero outputs. Figure~\ref{fig:crosstalk_example}(c) shows the counts of zero and non-zero outputs for these circuits over a hundred trials. Circuit-a, which avoids the crosstalk-prone link between Q2 and Q3, shows the highest count of zero outputs. The other two circuits, using links susceptible to crosstalk, exhibit a significant decrease in zero counts. These circuits are based on previous studies demonstrating both idling and operational crosstalk on IBM quantum hardware~\cite{das2021adapt}.   

\begin{figure*}[t]

\centering

\begin{subfigure}[b]{0.32\textwidth}
    \includegraphics[width=1\textwidth]{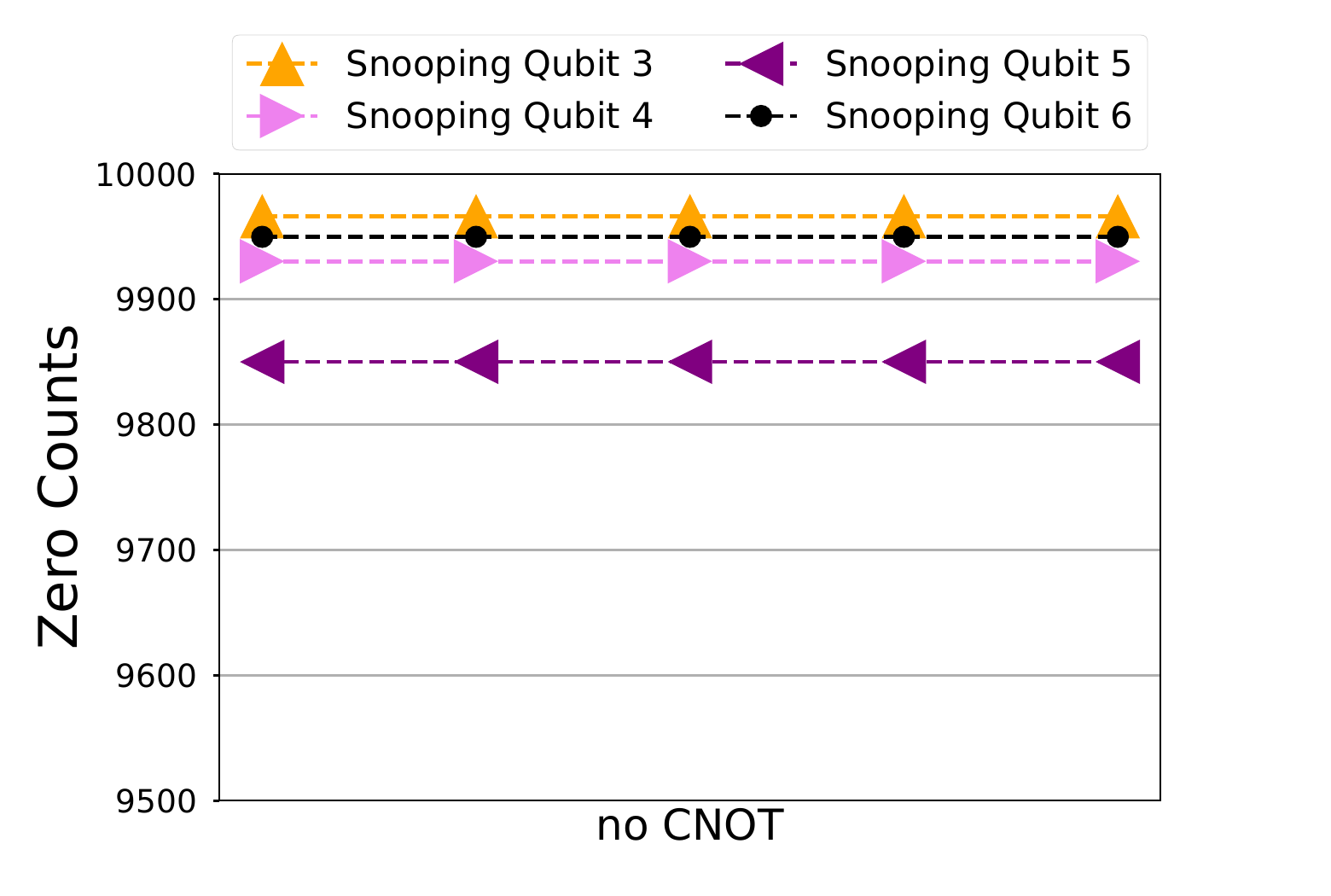}
    \caption{}

\end{subfigure}
\begin{subfigure}[b]{0.32\textwidth}
    \includegraphics[width=1\textwidth]{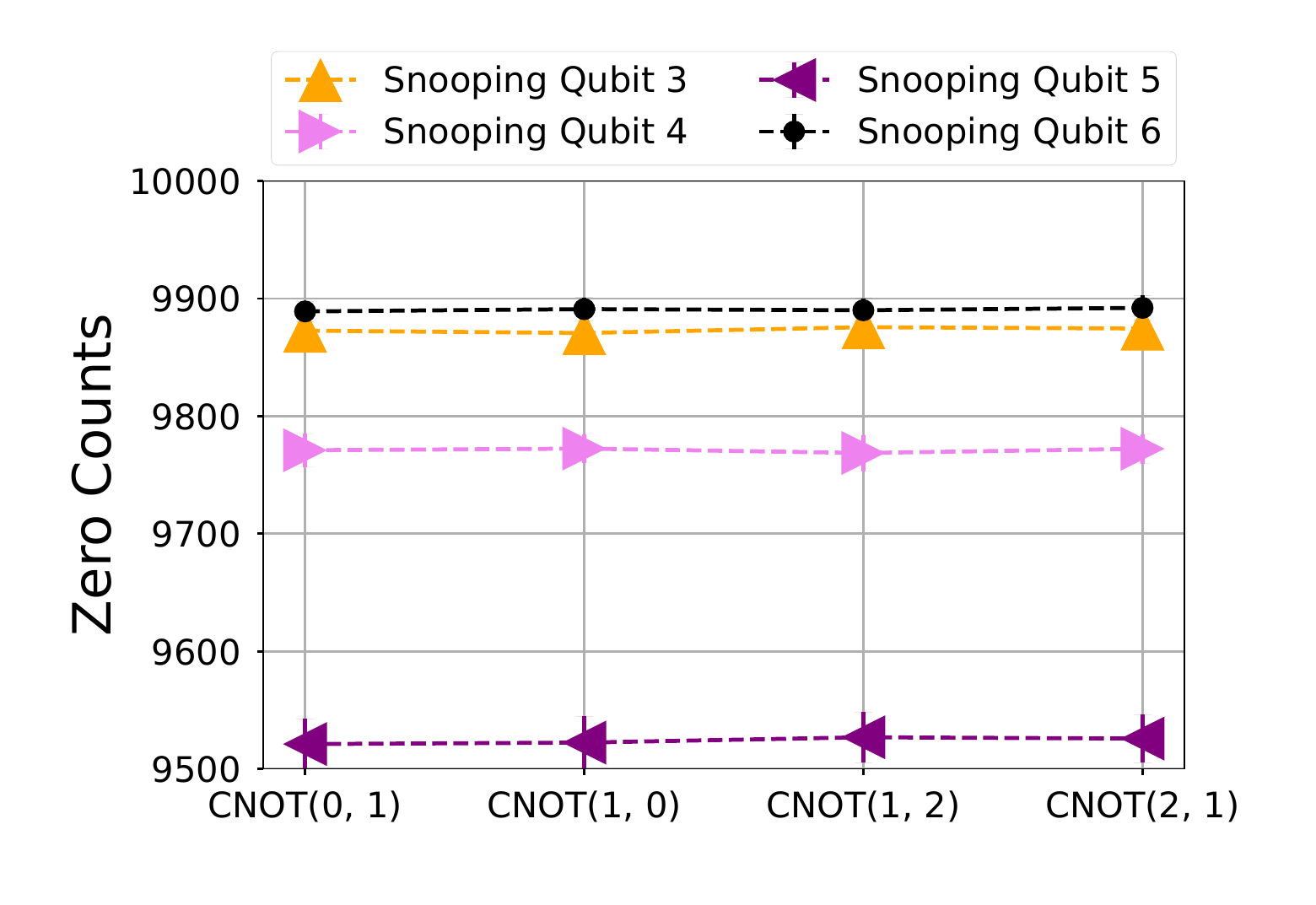}
    \caption{}

\end{subfigure}
\begin{subfigure}[b]{0.32\textwidth}
    \includegraphics[width=1\textwidth]{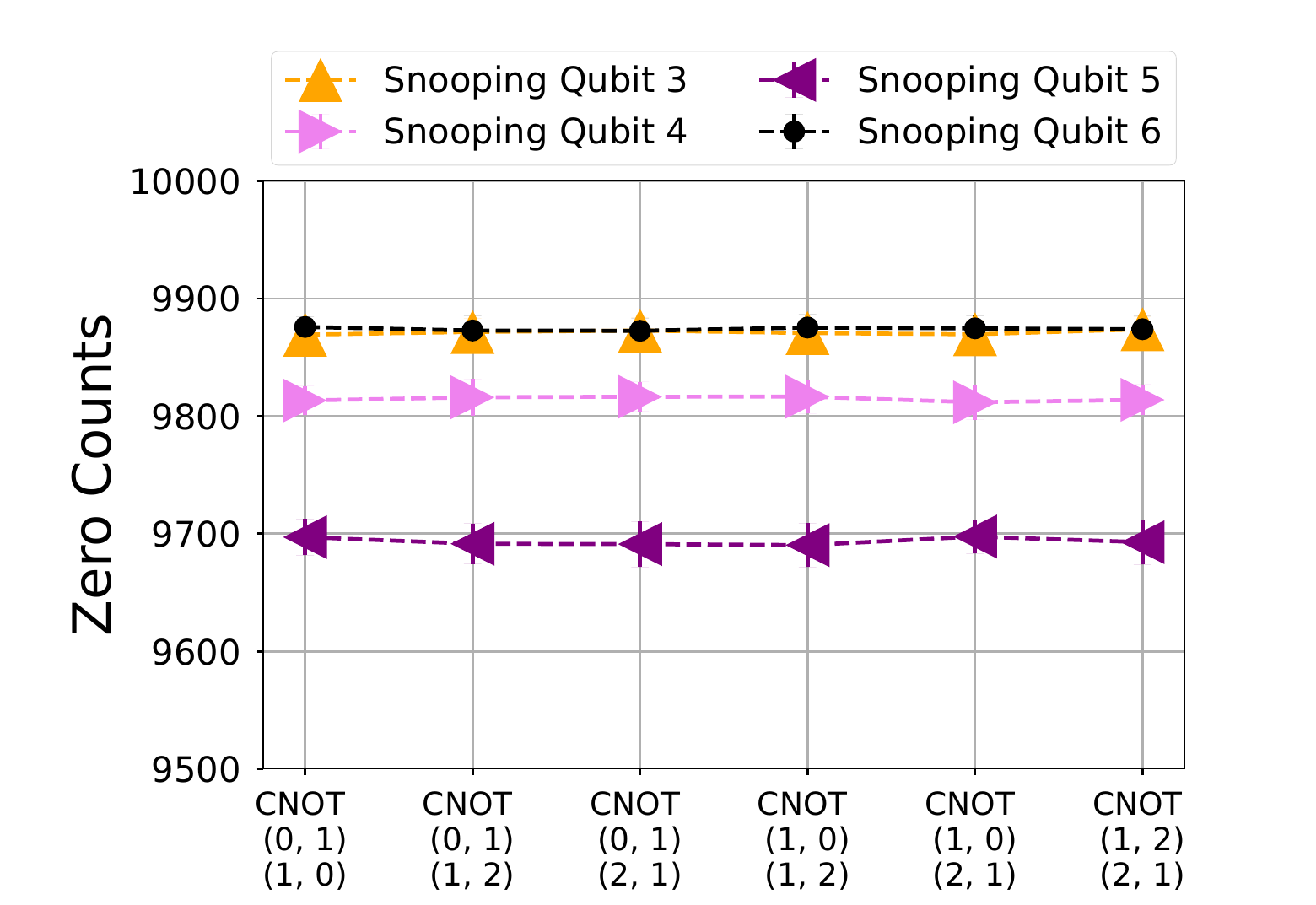}
    \caption{}

\end{subfigure}
\vspace{-0.1in}
\caption{Output of GCD circuits executed for a total of 10,000 iterations running on snooping qubits 3,4,5,6 of \texttt{IBM Lagos}, while victim qubits 0, 1, 2 runs (a) no CNOT gate, (b) one CNOT gate, (c) two CNOT gates.}\label{fig:crosstalk_detection}
\vspace{-0.2in}
\end{figure*}

\subsection{A Blueprint for Side Channel Attack} \label{subsec:sca}

Multi-tenancy is enabled by partitioning quantum hardware and assigning a subset of qubits to individual users. In such a setting, the attacker can exploit crosstalk to infer sensitive information. We use the following terminologies henceforth:  
\begin{itemize}
    \item \textit{Victim qubits} belong to the user that the adversary is targeting. Per our threat model, the adversary has no control over victim qubits.   
    \item \textit{Adversary or snooping qubits} are the set of physical qubit devices on quantum hardware that the attacker has control over, $i.e.$, the attacker can perform gates, measure, and analyze these qubits as much as any other user that has access to the quantum system.
    \end{itemize}
An attacker can detect crosstalk by continuously running a crosstalk detection circuit on snooping qubits. On most superconducting quantum computers, a two-qubit gate, such as CNOT generates a significant amount of crosstalk.
By observing this crosstalk, the attacker can deduce the count and position of the two-qubit gates in use. This information can reveal the program being executed by the victim.

To determine the number and position of two-qubit gates, we employ circuits like circuit-C in Figure~\ref{fig:crosstalk_example} (a). In these circuits, the idling error on the snooping qubit (Q1) is magnified by crosstalk from a CNOT operation on the victim qubits (Q2 and Q3). This effect is identifiable by measuring the zero counts. It's worth noting that while crosstalk from simultaneous gate operations can also be used for this purpose, such an approach requires the attacker to execute two-qubit gates. These gates are susceptible to errors unrelated to crosstalk, leading to reduced accuracy in detecting two-qubit gates.

\subsection{Side Channel on IBMQ Hardware} \label{subsec:scademo}
This section showcases \textit{Crosstalk-based Gate Detector (CGD)} protocol. To the best of our knowledge, this is the first demonstration of a side channel leakage exploiting crosstalk on real quantum hardware. 


\paragraph{Experimental Setup} To understand the effects of these possible information leaks in real systems, we used two IBM quantum computers with seven qubits: \texttt{IBM Lagos} and \texttt{IBM Perth}. To understand the efficacy of our attack, we permute attack and victim qubits, and to ensure statistical rigor, we repeat each attack experiment at least 75 times.  

Figure~\ref{fig:crosstalk_setup}~(a) shows the qubit connectivity map. To demonstrate the side channel, we use Q0, Q1, and Q2  as victim qubits, and the rest are used as attack qubits. Our goal is to detect, count, and localize CNOT gates running on the victim qubits using attack qubits.  We run a circuit shown in Figure~\ref{fig:crosstalk_setup}~(b), which creates an equal superposition of all the attack qubits, generating a state that is highly sensitive to crosstalk\footnote{Crosstalk often manifest as phase noise leading to local phase errors e.g. $\alpha\ket{0}+\beta\ket{1}$ $\xrightarrow[]{crosstalk} $$\alpha\ket{0}+ e^ {-i\phi}\beta\ket{1}$ }. This circuit is executed  10,000 times. The four snooping qubits are kept idling for a 400 nanosecond time window (typical CNOT gate duration on IBM hardware) and then measured. If there is no crosstalk, upon measuring the snooping qubits, all zero states (denoting the states of all qubits at `0') should appear with a very high probability. With crosstalk, all zero-state probability is reduced, and a crosstalk error generates non-zero states. 

Figures~\ref{fig:crosstalk_detection}~(a), (b), and (c)  show the zero counts for snooping qubits Q3, Q4, Q5, and Q6 when no (or zero) CNOT gate,  one CNOT gate, and two CNOT gates were applied on the victim qubits, respectively. Note that for all the experiments, we idle the snooping qubits in an equal superposition state for a fixed amount of time; this is done to ensure that we are mimicking a realistic attack scenario (as delineated in Section \ref{sec:threat}) where the attacker has no prior knowledge of CNOT counts.     





Moreover, each experiment is repeated 75 times to ensure that the observed data is statistically significant. In these repeated experiments, we observe a minimal variance in zero counts. Figures~\ref{fig:crosstalk_detection} (a), (b), and (c) show tiny error bars for all pairs of snooping qubits and CNOT gates. The results show a high all-zero count for the snoop qubits when no CNOT gates are used, as depicted in Figure~\ref{fig:crosstalk_detection} (a). However, applying one or more CNOT gates to the victim qubits reduces the zero count, indicating errors caused by crosstalk, as seen in Figures~\ref{fig:crosstalk_detection} (b) and (c). It is important to note that the impact of crosstalk varies among snoop qubits, dependent on factors like the device resonance frequency and specific spurious couplings between qubit pairs. We have observed similar patterns on \texttt{IBM Perth}, a seven qubit machine similar to \texttt{IBM Lagos}, and \texttt{IBM Guadalupe}, which is a sixteen qubit machine.

\begin{figure}[b]
\vspace{-0.2in}
    \centering
    \includegraphics[width=0.8\linewidth]{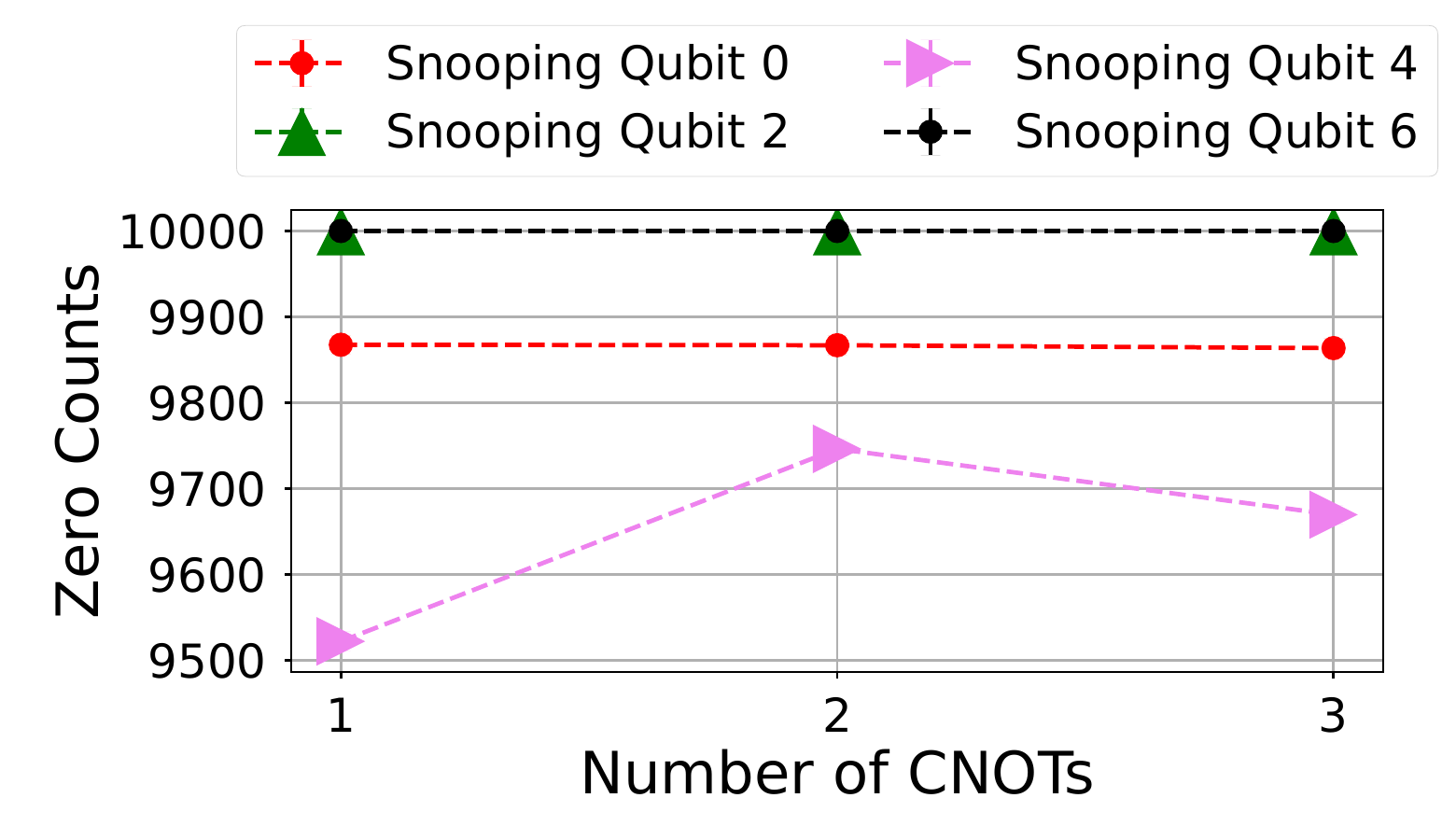}
    \caption{Distinguishability of number of CNOTs.}
    \label{fig:varying CNOTs}
\vspace{-0.2in}
\end{figure}

The number of zero counts does not always reduce as we apply more CNOT gates. Instead, these counts fluctuate because crosstalk mainly causes phase errors, which fluctuate. Figure~\ref{fig:varying CNOTs} demonstrates how zero counts change with different numbers of CNOT gates used on various combinations of victim qubits. We identify three distinct behaviors in qubit devices: first, there are qubits like qubit-2 and qubit-6 that show no change in response to crosstalk, making them unsuitable for detecting the number of CNOT gates applied. Second, there are qubits that can only distinguish between the presence and absence of CNOT gates. Finally, there are highly sensitive qubits, like qubit-4, that provide clear and distinct counts for different numbers of CNOTs applied.



\begin{figure}[t]
\vspace{-0.1in}
    \centering
    \includegraphics[width=0.8\linewidth]{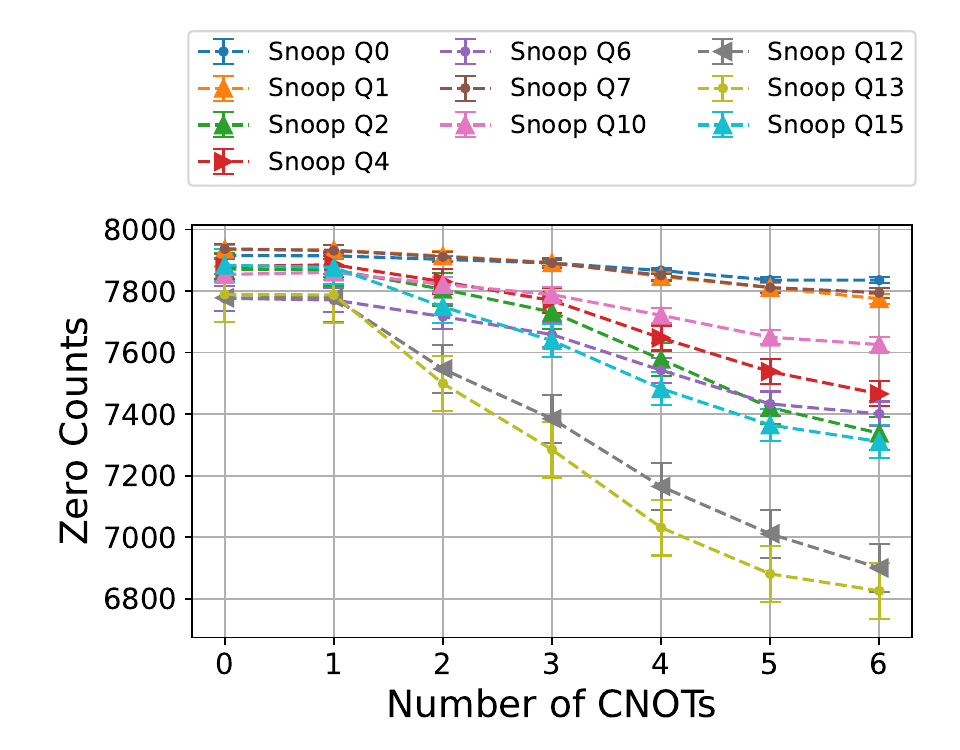}
    \vspace{-0.15in}\caption{Comparing the distinguishability of number of CNOTs based on zero counts on \texttt{IBM Guadalupe}. }
    \label{fig:guadalupe_data}
    \vspace{-0.2in}
\end{figure}

\subsection{Scalability of CNOT counting} \label{subsec:cnotcount}
To evaluate if our method can be applied more widely, we created random quantum circuits with varying number of CNOT gates affecting the victim qubits. Meanwhile, the snooping qubits were set to an equal superposition state and then left idle for a consistent period. These experiments were conducted on a 16-qubit IBM quantum computer. Figure~\ref{fig:guadalupe_data} illustrates how the zero counts correlate with the number of CNOT gates used. We observe a strong negative correlation between zero counts and the number of CNOTs for many snooping qubits. Note that the CNOT counts are averaged over many random circuits running on different permutations of victim qubits with fixed CNOT counts. Our experiments demonstrate a possibility that the attacker can run CGD and use the zero counts to infer the CNOTs in the victim's circuits. 


Although dominant and ubiquitous, crosstalk can be suppressed using software techniques such as dynamical decoupling, where a sequence of crosstalk suppression operations is continuously run on the qubits~\cite{volume64}. We investigated how well our side-channel attack method would work if a quantum cloud provider used Dynamical Decoupling (DD) to prevent crosstalk. With a protective DD setting, an attacker couldn't leave their qubit idle to detect crosstalk, as the cloud service provider would be applying decoupling operations in the background to block any potential side-channel threats. On \texttt{IBM Lagos}, we observe that despite applying DD, we could still see almost identical zero counts similar to Figure~\ref{fig:varying CNOTs}. Moreover, as the number of qubits on the quantum computers increases, crosstalk between pairs of qubits increases as well, leading to more prominent side channels.

\subsection{Desensitizing qubits against other errors}
It is important to note that in addition to crosstalk errors, readout errors are another significant source of inaccuracies in the output of quantum circuits. To mitigate the misdiagnosis of readout errors as crosstalk errors, we initially set the basis of the qubits in the adversary's circuit to the \(\ket{+}\) state and revert them to the \(\ket{0}\) state before measurement, as discussed in Section \ref{subsec:cross}. Since our interest lies in measuring deviations from zero, the impact of readout errors is minimal. This is because a transition from the \(\ket{0}\) to the \(\ket{1}\) state involves qubit excitation, which occurs much less frequently than the relaxation from \(\ket{1}\) to \(\ket{0}\). Moreover, existing methods like flip-and-measure remove the bias of readout errors, thereby eliminating their interference in the crosstalk readings we obtain~\cite{tannu2019mitigating}. Furthermore, as demonstrated in Section \ref{subsec:fuzz}, the minimal degree of readout errors, aside from crosstalk errors, does not significantly affect the overall performance of our circuit identification model.
\section{Proposed Attack Methodology} \label{sec:gnn}

\begin{figure}[b]
\vspace{-4mm}
\centering
  \includegraphics[width=\linewidth]{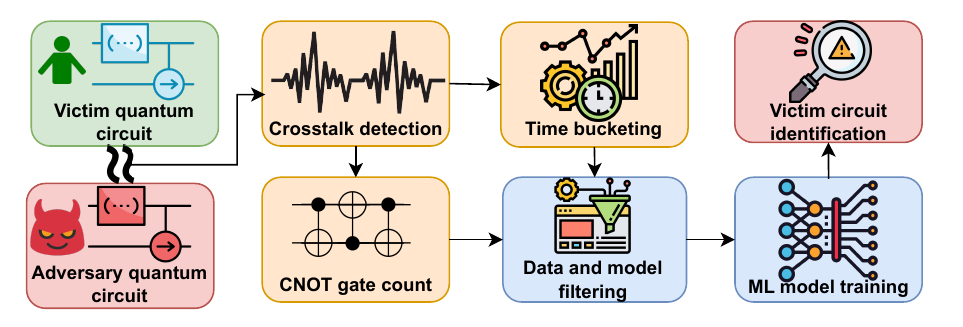}
  \caption{Overview of proposed quantum side-channel analysis for victim circuit identification.}
  \label{fig:atck_surface}
  \vspace{-4mm}
\end{figure}

In this section, we delineate our proposed attack, which uses a quantum side channel exploiting prevalent crosstalk in QaaS devices. To this end, relevant data is initially extracted from the crosstalk signature, following which an appropriate classification model is selected to learn the intricate insights from this data, as detailed in Sections \textbf{\ref{subsec:attack_surface}} and \textbf{\ref{sec:graph_generation}}, respectively. Subsequently, the selected classifier is trained with relevant features necessary for identifying the circuit being executed by the victim, as described in Section \textbf{\ref{sec:GNN_training}}.

\subsection{Attack Surface} \label{subsec:attack_surface}


In order to launch an attack adhering to our threat model (Section \ref{sec:threat}), we first analyze the spatial and temporal distribution of the CNOT gates present in the victim quantum circuit following the extraction of crosstalk signals from it (as presented in Section \textbf{\ref{subsec:scademo}}). The goal of this analysis is to extract critical information pertaining to the victim circuit which can potentially enable its identification. An overview of the procedural steps in executing our side channel attack is depicted in Figure \ref{fig:atck_surface}. The figure offers a comprehensive depiction of the constituent elements comprising the attack surface, which are explained as follows:
\begin{enumerate}

    \item \underline{\textbf{\textit{Strategic allocation of adversarial qubits}}}: The snooping qubits are strategically positioned in the quantum computer in a manner enabling crosstalk detection from the victim qubits within the physical architecture of the quantum computer being employed (as mentioned in Section \ref{subsec:sca}).
    \item \underline{\textbf{\textit{Crosstalk detection}}}: The snooping qubits are sensitized by changing their basis to detect the crosstalk signature from the victim circuit (as delineated in Section \ref{subsec:scademo}). 
    \item \underline{\textbf{\textit{CNOT gate count}}}: Based on the crosstalk signature detected by snooping qubits from the affected zero counts, the total number of CNOT gates present in the victim circuit can be determined (mentioned in Section \ref{subsec:cnotcount}). 
    \item \underline{\textbf{\textit{Time bucketing of CNOT operations}}}: Following the determination of the total number of CNOT gates in the victim circuit, the crosstalk signature is divided into time buckets. The primary objective of this time bucketing is to obtain the time-localized CNOT connectivity in the victim circuit. An illustration of time bucketing is depicted in Figure \ref{fig:time_div}, where the overall operational schedule of a four-qubit Variational Quantum Eigensolver (\textit{VQE}) is divided into four time buckets with uniform duration \textbf{\textit{tb}}. This division is conducted following the selection of a time bucket duration, and executing the adversarial quantum circuit on the snooping qubits with distinct sensitized durations, with the help of delay gates. This enables crosstalk inference from different victim circuit segments.
    
    Figure \ref{fig:time_div} shows the number of CNOT gates connecting qubit pairs in each time bucket, which reveals the temporal connectivity of victim circuit qubits facilitated by CNOT gates. 
    This approach increases the likelihood of discerning the structural characteristics of the victim circuit by enhancing the performance of our GNN-based classification model (discussed in Section \ref{sec:results}).

\begin{figure*}[t]
\centering
 \begin{subfigure}[b]{0.24\linewidth}
 \includegraphics[width=\textwidth]{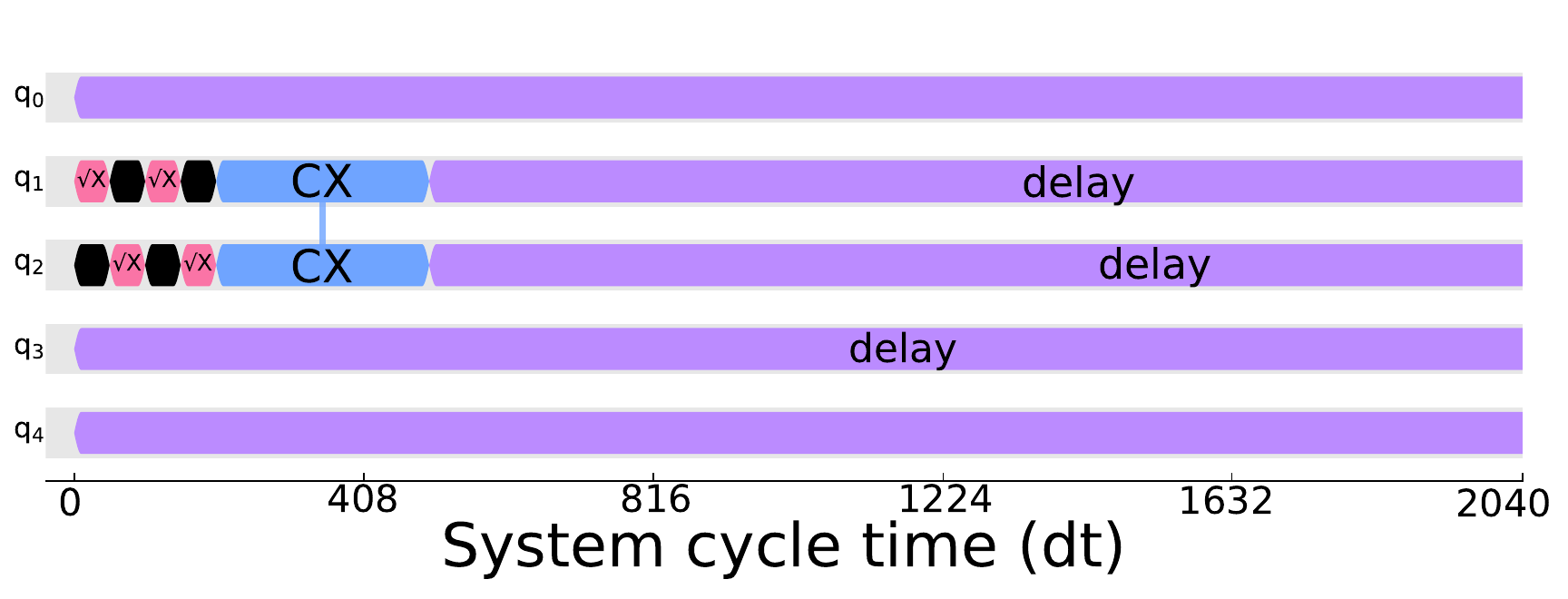}
 \caption{Time bucket 1 containing \textit{one} CNOT gate.}
 \label{t1}
 \end{subfigure}
 \begin{subfigure}[b]{0.24\linewidth}
 \includegraphics[width=\textwidth]{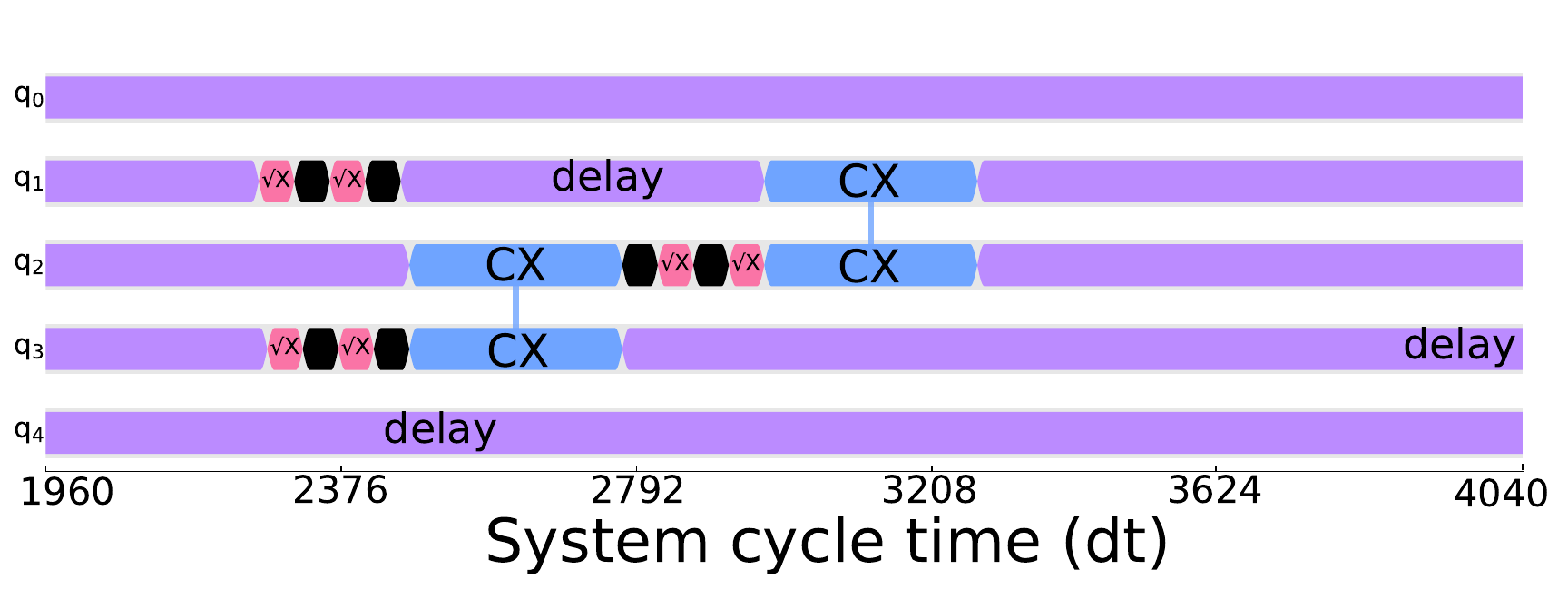}
 \caption{Time bucket 2 containing \textit{two} CNOT gates.}
 \label{t2}
 \end{subfigure}
 \begin{subfigure}[b]{0.24\linewidth}
 \includegraphics[width=\textwidth]{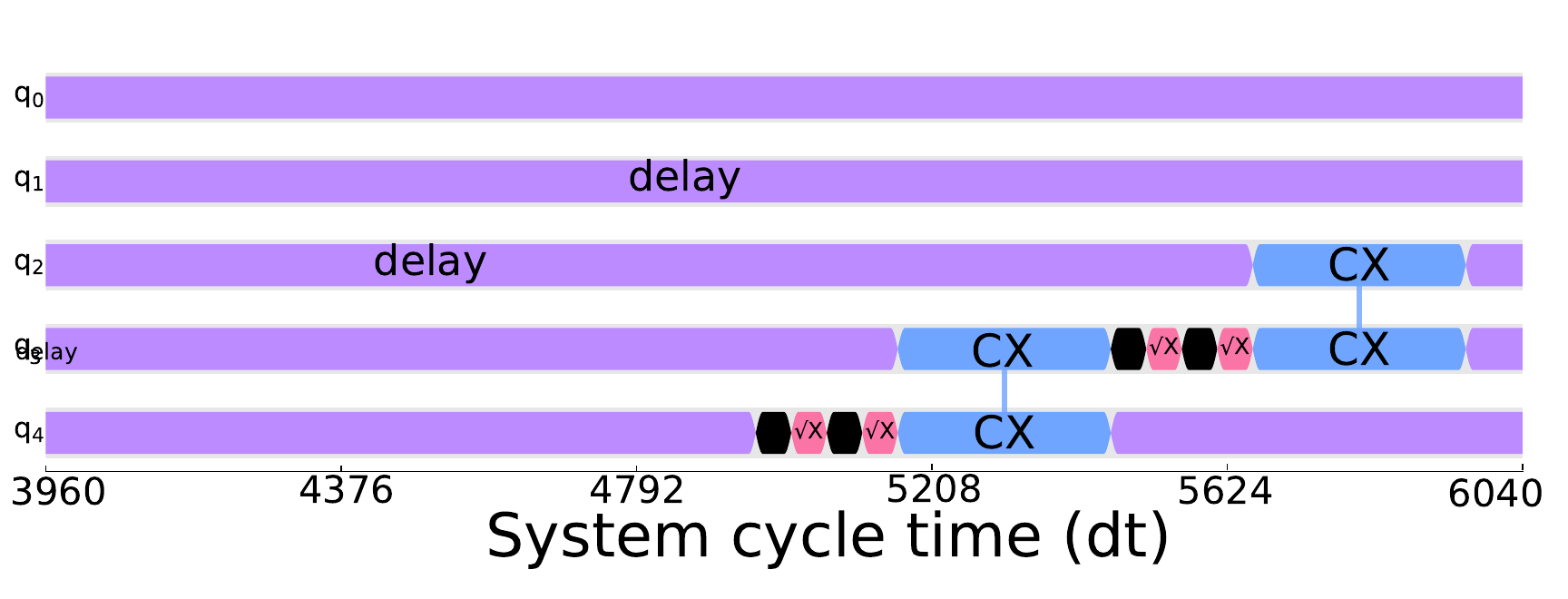}
\caption{Time bucket 3 containing \textit{two} CNOT gates.}
\label{t3}
 \end{subfigure}
 \begin{subfigure}[b]{0.24\linewidth}
 \includegraphics[width=\textwidth]{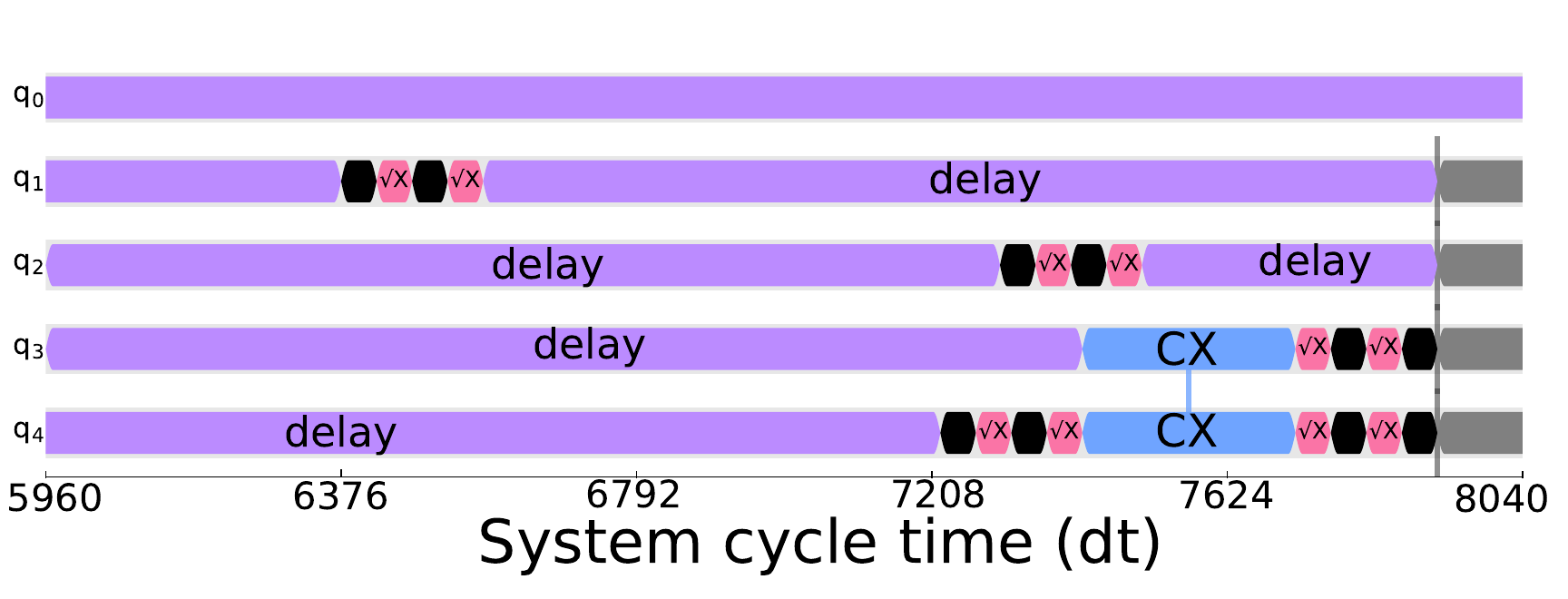}
 \caption{Time bucket 4 containing \textit{one} CNOT gate.}
 \label{t4}
 \end{subfigure}
 \vspace{-2mm}
 \caption{Example of the time bucketing operation conducted on a four-qubit VQE circuit. With a uniform time bucket duration of \textbf{\textit{tb}} = 2000 \textbf{\textit{dt}} (where \textbf{\textit{dt}} is the system cycle time, equivalent to \textit{2.22} \textbf{ns}), the adversarial circuit is divided into four time buckets, and temporal CNOT-gate distribution in the victim circuit is obtained, as displayed in Figures \ref{t1} to \ref{t4}.}
 \label{fig:time_div}
 \vspace{-6mm}
\end{figure*}

\item \underline{\textbf{\textit{ Robust detection using representative circuit datasets}}}: \label{subsubsec:class}
Following the successful acquisition of the position and count of CNOT gates present in the victim circuit, this data is further streamlined to facilitate the identification of the victim circuit. 
It is worth noting that the complexity of this analysis is non-trivial since each quantum circuit can encompass a total of 128 potential variations, which arise due to quantum circuit transpilation prior to their execution. Transpilation is the process of optimizing and transforming a quantum circuit into an equivalent circuit that can be efficiently executed on specific quantum hardware while respecting its physical constraints and gate set~\cite{younis2022quantum}.
There are four stages of transpilation, each of which has a subset of implementations. Prevalent QaaS vendors offer users the capacity to select one implementation for each stage, namely \textbf{\textit{four}} distinct layout stage methods, \textbf{\textit{four}} distinct routing stage methods, \textbf{\textit{four}} different optimization stage levels, and \textbf{\textit{two}} scheduling stage methods~\cite {IBMQDocumentation}. 
To streamline our analysis in a practical manner, we make the following observations. 
\begin{enumerate}
    \item Firstly, given that transpilation grants user control, it is reasonable to presume that users seek a high-fidelity output from their executed circuits~\cite{cross2019validating}. Consequently, it is justified to assume only the best optimization level is applied to the victim circuit.
    \item Next, users will generally select their scheduling method in pursuit of higher output fidelity to minimize errors due to decoherence~\cite{smith2021error}. Consequently, it is fair to assume that victim circuits will be executed using the \textit{`as-late-as-possible'} scheduling method, the application of which furnishes higher output fidelity compared to the alternative execution of \textit{`as-soon-as-possible'}.
\end{enumerate} 

Since optimization levels and scheduling methods beyond the ones chosen are predominantly utilized for experimental purposes rather than having widespread commercial or scientific applications, we select these transpilation configurations for our approach.
Consequently, the total number of possible transpiled circuits for each benchmark is effectively reduced by eight times (by selecting transpilation configurations with the best optimization level out of four, and the best scheduling method out of two).

To orchestrate a successful attack, it is necessary to identify, extract, and learn fundamental information contained in the structure of the quantum circuits. With the ultimate goal of successfully classifying the circuit as one of the potential victim circuits (which are limited, as mentioned in Section \ref{sec:threat}), we make three observations related to the available information extracted from the acquired data. 
\begin{itemize}
\item \noindent\textbf{Observation 1:} The aggregate count of CNOT gates in a victim circuit is available to us. 
To highlight its relevance in circuit distinction, we studied their variance across 10 distinct benchmarks from the \textit{scalable} category in the widely utilized MQTBench, and the findings are shown in Table \ref{tab:bench}. Column one in the table highlights the benchmarks, followed by column two which depicts the number of CNOT gates in the circuit. Column three denotes the average CNOT gates per qubit pair, and finally, column four highlights the average number of CNOT gates present in a given time bucket (duration was taken to be 3000 $dt$).
This table highlights the range of variation in the number of CNOT gates across the different 120-qubit benchmark circuits, from 119 to 30576 CNOT gates across benchmarks. This can be a relevant feature since CNOT gates are responsible for connections among qubits in a quantum circuit.
\item \noindent\textbf{Observation 2:} In addition to the total number of CNOT gates, the \textit{number of CNOT gates per qubit pair} is also available. This is obtained by using distinct snooping qubits to extract crosstalk from different CNOT gates by analyzing the zero counts, as discussed in Section \ref{sec:xtalk}. This captures the intricate qubit state changes caused by CNOT operations and can be encoded as a key feature for our classification model, providing valuable insights into the circuit's structure.
\item \noindent\textbf{Observation 3:} Finally, the temporal sequencing of the involved CNOT gates is available to us. This is of high importance since by understanding the temporal aspects, it is possible to gain insights into the precise moments when these gates influence the quantum states of qubits, contributing to a more nuanced comprehension and thereby enabling circuit identification.
\end{itemize}


\begin{table}[t]
\centering
\caption{Variance in number of CNOT gates across all \textit{scalable} \textbf{120-qubit} benchmark circuits from MQTBench~\cite{quetschlich2023mqtbench}. The x-axis represents the benchmark numbers taken in alphabetical order as they appear in the MQTBench suite, and the y-axis represents the number of CNOT gates.}
    \label{tab:bench}
\resizebox{\linewidth}{!}{%
\begin{tabular}{c|c|c|c}
\hline

\textbf{Benchmarks}        & \textbf{CNOT gate count} & \textbf{\begin{tabular}[c]{@{}c@{}}Avg. CNOT gates \\ between qubit pairs\end{tabular}} & \textbf{\begin{tabular}[c]{@{}c@{}}Avg. concurrent CNOT gates\\ (in a time bucket)\end{tabular}} \\ \hline \hline
ae\_nativegates\_n120            & 14280                      & 2                                                                                      & 7                                                                                                \\ \hline
dj\_nativegates\_n120            & 119                      & 1.1                                                                                       & 2                                                                                                \\ \hline
ghz\_nativegates\_n120           & 119                      & 1                                                                                       & 2                                                                                                \\ \hline
graphstate\_nativegates\_n120    & 120                      & 1.3                                                                                       & 3                                                                                                \\ \hline
qft\_nativegates\_n120           & 14460                    & 2.03                                                                                      & 9                                                                                               \\ \hline
qnn\_nativegates\_n120           & 28679                      & 4.02                                                                                      & 14                                                                                                \\ \hline
qpeexact\_nativegates\_n120      & 14457                     & 2.5                                                                                      & 12                                                                                               \\ \hline
random\_nativegates\_n120        & 30576                    & 4.93                                                                                      & 22                                                                                               \\ \hline
realamprandom\_nativegates\_n120 & 21420                    & 3.3                                                                                      & 16                                                                                               \\ \hline
se2random\_nativegates\_n120     & 21422                    & 3                                                                                      & 11                                                                                               \\ \hline
\end{tabular}%
}
\vspace{-6mm}
\end{table}

\begin{figure*}[t]
 \raggedright
 \begin{subfigure}[b]{0.44\textwidth}
 \includegraphics[width=\textwidth]{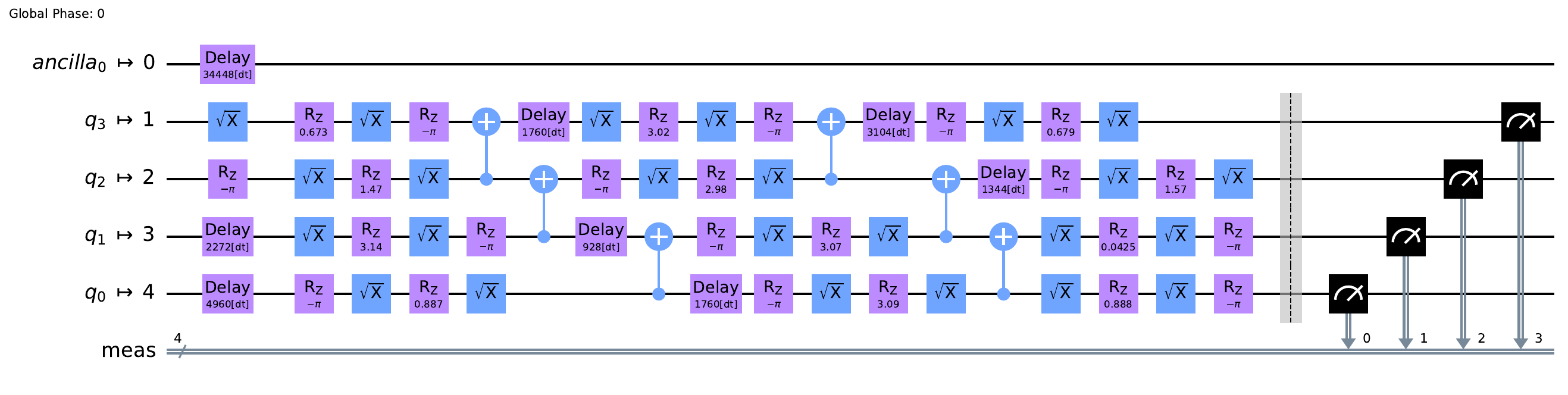}
 \caption{Transpiled circuit obtained following the transpilation passes, which will be executed. }
 \label{taddet}
 \end{subfigure}
 \begin{subfigure}[b]{0.3\textwidth}
 \includegraphics[width=\textwidth]{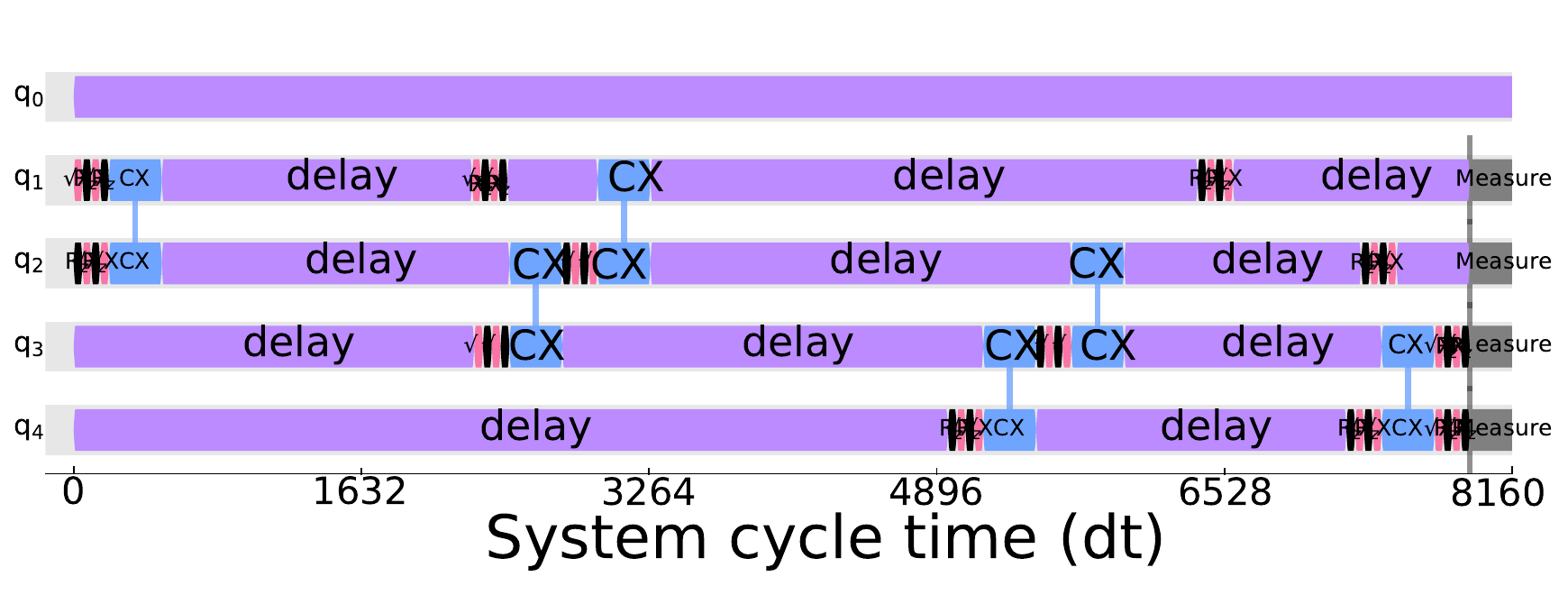}
 \vspace{-4.5mm}
 \caption{Operational schedule of the transpiled circuit during execution.}
 \label{timingadder}
 \end{subfigure}
 \begin{subfigure}[b]{0.24\textwidth}
 \centering
 \includegraphics[width=\textwidth]{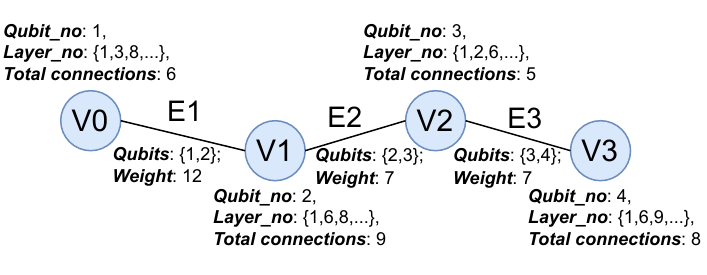}
 \caption{Graph representation of the scheduled transpiled circuit.}
 \label{graphadder}
 \end{subfigure}
 \caption{This figure demonstrates the conversion of data obtained from a transpiled quantum circuit and its operational schedule. Figure \ref{taddet} depicts a transpiled four-qubit VQE circuit mapped to a five-qubit IBMQ Lima. Figure \ref{timingadder} demonstrates the temporal distribution of the quantum gates in the transpiled circuit as they will be executed. Figure \ref{graphadder} provides an example of the graph encoding by representing qubits as nodes (\textit{V}), and CNOT gates as edges (\textit{E}), thereby forming the graph with information encoded in nodes and edges: \textit{G = (V,E)}. The relevant node and edge features extracted are also depicted here. }
 \label{fig:rq2_3}
 \vspace{-6mm}
 \end{figure*}

\item \underline{\textbf{\textit{Data-informed ML classifier selection}}}:\label{section:classifier_selection}
Subsequent to these observations, it is worth noting that the salient information available for training the classification model can be separated into two distinct components. The first component comprises information pertaining to the involved qubits, followed by the second component, encompassing the data derived from the interconnections among these qubits. 
This essentially ensues from the fact that any quantum circuit resembles a graphical structure, with the qubits as nodes, and CNOT connections between the qubits resembling edges between the nodes in a graph~\cite{pmlr-v97-zhang19e}.
Consequently, given the inherent structure of this data and guided by our observations, it is discerned that the data extracted from the victim circuit can be effectively represented in the form of a graph~\cite{He2023, 9877896}. This realization has, in turn, prompted the application of a Graph Convolutional Network (GCN)-based classifier for identifying the victim circuit accurately. The selection of our GCN-based classification model is further substantiated by its suitability to handle graph-structured data and its adeptness at analyzing classical circuits~\cite{Said2023, dong2023cktgnn}. 

\end{enumerate}

\subsection{Graph Generation and Feature Extraction}\label{sec:graph_generation}

For developing the GCN model, each qubit involved in computation is considered a node, and each CNOT gate connecting two qubits in the circuit is considered an edge. 
The graph node and edge features are defined as follows:

\begin{enumerate}
    \item \textbf{Qubit number:} The number of qubits that are involved in the quantum circuit execution can be obtained experimentally, as demonstrated in Section \ref{subsec:scademo}. For example, the VQE circuit demonstrated in Figure \ref{fig:rq2_3} comprises four qubits, each of which we consider a node.
    \item \textbf{Total CNOT gates on each qubit:} The total count of edges (encompassing both cases where the node is \textit{control qubit}, or a \textit{target qubit}) linked to a node is designated as the number of connections. The inclusion of this attribute follows from observations 1 and 2 outlined in Section \ref{section:classifier_selection}. To illustrate this, in figure \ref{fig:rq2_3}, the transpiled circuit comprises two CNOT gates on qubits 1 and 4, and four CNOT gates on qubits 2 and 3. 
    \item \textbf{Time distribution of CNOT gates:} The temporal distribution of CNOT gates is obtained by dividing the quantum circuit execution into time buckets with a duration of \textit{$\mathbf{\textit{tb}}$}. This attribute corresponds to observation 3 highlighted in Section \ref{section:classifier_selection}. 
    The minimum value of \textit{$\mathbf{\textit{tb}}$} is equivalent to the duration of execution of a single CNOT gate. This consideration arises from the fact that if \textit{$\mathbf{\textit{tb}}$} is smaller than the duration of a single CNOT gate, detecting the presence of CNOT gates would not be possible. For example, from the schedule of the VQE circuit, with \textit{$\mathbf{\textit{tb}} = 2000dt$}, the time distribution is shown in Figure \ref{fig:time_div}.
    \item \textbf{Active qubits:} The concept of active qubits denotes the qubits engaged in CNOT gate operations, delineating the interconnections between them. This essential attribute helps in establishing a graphical representation of the quantum circuit. For example, in Figure \ref{fig:rq2_3}, the physical qubits 1,2,3, and 4 are involved in the transpiled circuit, while qubit 0 remains idle for the duration of computation.
    \item \textbf{Number of CNOT gates per qubit pair:} The number of CNOT gates per qubit pair is represented as the weight of each edge signifies the number of CNOT gates present between the qubits involved.

\end{enumerate}
The structural information captured by these features is crucial for identifying the victim quantum circuit using the GNN classifier, as shown in Figure \ref{fig:rq2_3} using a four-qubit half-adder circuit. To build a robust training dataset, we extract fine-grained qubit connectivity from each circuit to construct graph representations. Based on our prior knowledge of circuit types, we label each graph accordingly, defining distinct classes within the dataset.

\subsection{GCN Model Training}\label{sec:GNN_training}
This section describes how we build and optimize the GCN model for identifying the victim quantum circuit by assimilating the features outlined thus far.
\subsubsection{Model construction}
The graph built in Section \ref{sec:graph_generation}, accompanied by the node features, and labels, serves as the basis for training a GCN model, consisting of $G_l$ graph convolution
layers. 
The input layer has $F$ neurons,
where $F$ is the feature dimensionality of the graph nodes.
We define the model
as a function $GCN(x)$ producing output $y \in R^n$ for
input $x \in R^N$. The output of
$GCN(x)$ is a $n$-dimensional vector containing the probabilities
$P_c$ of classes ($C \in \{0,n-1\}$). 
 The
final prediction class $y_c$ is obtained by applying the $argmax$
function: $argmax(GCN(x)) = y_c$.

\begin{algorithm}[t]
        \flushright{
            \caption{GCN training for circuit identification} 
            \label{alg:5} 
            
            \begin{algorithmic}[1] 
            \REQUIRE ~~ 
            Quantum circuit Design $\mathcal{D}$;\\
            \ENSURE ~~
            Quantum circuit label $Y_c$;\\
            \STATE $G(V,E) \gets GetGraph(D)$
            \STATE ${A}_{N \times N} \xleftarrow{} AdjacencyMatrix(E)$
            \STATE $Features \xleftarrow{} ExtractFeatures(G,D)$
            \STATE ${X}_{ N \times F} \xleftarrow{} FeatureMatrix(Features)$
            
            \FOR{$ v \text{ in } V$ }
            \STATE ${e}_{v} \xleftarrow{} {E_f}(v)$
            \ENDFOR
            \FOR {$l$ in layers $L$ }
            
            \STATE ${e}_{v} \xleftarrow{} AggregateNeighbours({e}_{v} ,{A}_{N \times N},{W}_{i})$
            
            \STATE ${e}_{v} \xleftarrow{} ActivationFunction({e}_{v})$
            \ENDFOR
            
            \STATE ${e}_{v} \xleftarrow{} AggregateNodeFeatures({e}_{v} ,{X}_{ N \times F},{W}_{f})$
            
            \STATE ${e}_{v} \xleftarrow{} ActivationFunction({e}_{v})$
            
            \STATE $Y_c \xleftarrow{} logSoftmax(linear({W}_{y}*{e}_{v}))$
            
            \end{algorithmic}
        }
\end{algorithm}

\subsubsection{Model training}
We train the GCN model using the circuit's graph representation, node features, edge connections, and the labeled dataset created in Section \ref{sec:graph_generation}. Training involves using a subset of the dataset, while the remaining graphs are reserved for validation post-training. 
The GNN training process is detailed in Algorithm \ref{alg:5}, which utilizes the graph ($G$) generated from the design netlist ($D$) as the input argument. First, the edges 
and node features are extracted from the design (\textit{lines 2-3}). The extracted features are represented as a matrix of size \textit{N} $\times$ \textit{F}, where \textit{N} is the number of nodes and \textit{F} is the number of features (\textit{line 4}). Subsequently, the node embeddings for each node ($v$) is initialized using an embedding function $E_f$, which results in embeddings $e_v$ (\textit{lines 5-7}). Next, in each iteration of message passing, the embeddings of each node $v$ and that of its neighbors are aggregated using the matrix $A$ and a weight matrix $W_l$. The resulting vector is passed through an activation function, and the updated embedding $e_v$ is stored for the next iteration (\textit{lines 8-11}). Following the last iteration, the final node embeddings are computed by aggregating the original node features $X$, and a weight matrix $W_f$. These final embeddings can be subsequently utilized for node classification (\textit{lines 12-14}), where the class probabilities $Y_c$ are calculated by passing the final embedding $e_v$ through an activation function \textit{log\_softmax}. 
The class label can be obtained by applying $argmax$ function on the class probabilities. The class labels are the benchmark circuits, and accurate prediction of this label by our GCN model signifies successful identification of the victim circuit.

\section{Evaluation of proposed attack} \label{sec:results}

In this section, we evaluate the proposed attack framework's performance in successfully identifying the victim quantum circuit. To that end, we formulate a multi-class classification problem, wherein we classify the circuit executed by the victim using CNOT signatures detected by the adversary.  


\subsection{Experimental Setup} \label{subsec:expsetup}
\paragraph{\textbf{Benchmarks}} To evaluate our proposed attack, we consider a total of 336 benchmark circuits, which are obtained from MQTBench, with the maximum number of qubits being limited to 30. This limitation is imposed due to decoherence considerations, inhibiting the useful execution of large quantum circuits beyond 25 qubits~\cite{quetschlich2023mqtbench}. This resulted in a total of 336 benchmark circuits obtained. \textit{Additionally, each benchmark circuit is further divided into 16 transpiled circuits, as discussed in Section \ref{subsubsec:class}, furnishing a total of 5376 distinct circuits, for a comprehensive evaluation of our proposed attack model}.

\paragraph{\textbf{Evaluation metrics}} In the assessment of our GCN model's performance, we analyze the accuracy, precision, and recall furnished by our GCN classification model to determine its efficacy in the task of victim circuit identification. Subsequently, we present its performance concerning the selection of optimal model parameters, with the overarching objective of optimizing the overall model performance. We further detail the robustness of our model design by evaluating its performance in the presence of fuzziness or inaccuracies that can creep in during real-world data acquisition. 

\subsection{GCN framework evaluation} \label{subsec:GCNeval}


For victim circuit identification, we use a GCN classifier comprising three hidden layers, two fully connected layers, and a single ReLU layer.
For the 336 benchmark circuits, considering a time window of 2000 ns, the GCN model comprising 64 hidden channels averaged over 1000 iterations, 
furnishes an overall accuracy of 85.6\%, with a precision of 93.1\% and a recall value of 92.7\%. 
To further improve the performance of our model, we study its performance by varying the resolution of the dataset and changing the design parameters, as follows.


\subsubsection{Resolution Optimization}
\begin{figure}[t]
\centering
  \includegraphics[width=0.75\linewidth]{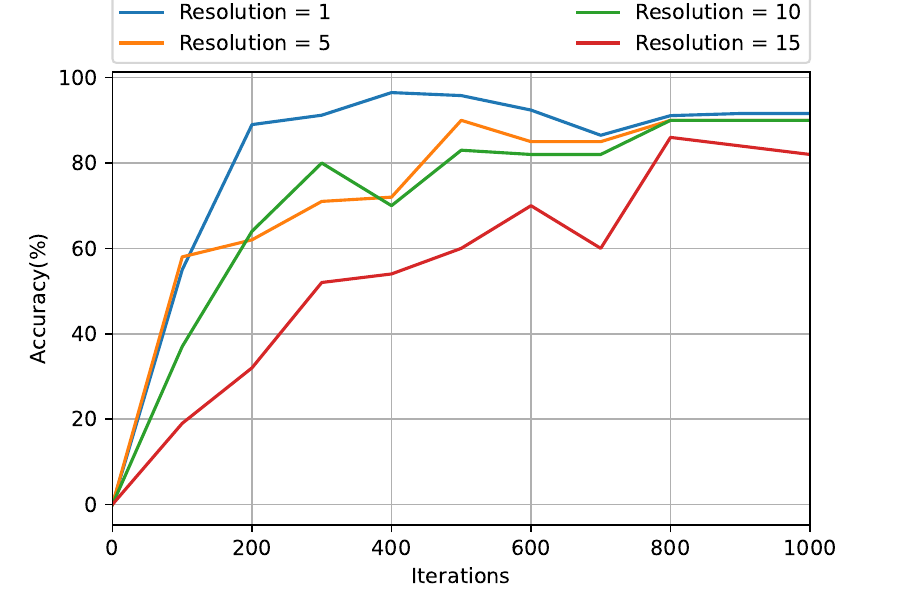}
  \caption{Average circuit detection accuracy of proposed GCN model under varying resolutions.}
  \label{fig:resolution}
  \vspace{-4mm}
\end{figure}

To enhance the performance of our model, the selection of an appropriate dataset 
is of paramount importance. Since the benchmark circuits are partitioned into time buckets (as mentioned in \ref{subsec:attack_surface}), it is necessary to determine the optimal duration of each time bucket which leads to enhanced classification accuracy for our GCN model.
Accordingly, we systematically vary the duration of each time bucket, thereby constructing multiple datasets, which we designate as the resolution of the dataset. Here, the term ``resolution" denotes the timing window duration employed in the creation of the dataset for our GCN-based model, as denoted in Equation \ref{eq:res}. 
    \begin{equation} \label{eq:res}
        Resolution = \frac{\text{$Count(CX_{Gate}$)} }{\text{$Duration(CX_{Gate})$}}
    \end{equation}
Equation \ref{eq:res} explicitly expresses that the resolution is equivalent to the ratio of the number of CNOT gates within a specified time window to the duration of a single CNOT gate. Following this formulation, a corresponding dataset is systematically generated, adhering to the temporal partition defined by the aforementioned resolution, and subsequently utilized as input for the GCN model.
The variation in accuracy furnished by utilizing datasets with different resolutions, spanning 1000 iterations, is demonstrated in Figure \ref{fig:resolution}. 
In this figure, the y-axis denotes the accuracy furnished by our GCN model, while the x-axis denotes the number of iterations over which the
resolution is varied in the range of 1-15. Here, a resolution of one corresponds to a time bucket duration same as the duration of execution of a single CNOT gate. 

From Figure \ref{fig:resolution}, it can be observed that for a resolution of 1, the accuracy furnished by the GCN model saturates at over 90\% with less than 200 iterations, compared to a resolution of 15, which achieves an accuracy of 70\% for over 1000 iterations. Thus, enhancing the resolution progressively increases the number of iterations required for our GCN model to make accurate predictions.
This phenomenon arises from the inherent presence of errors in real-world detection data obtained from experimental settings, which can notably influence the model's performance, especially when operating at very high resolutions. Consequently, such errors lead to an elevated occurrence of misclassifications, thereby impeding the classifier's effectiveness as the resolution increases.
Following our observations, for our GCN model, we utilize the dataset created with a resolution of one, which furnishes an accuracy of 85.6\% with a precision and recall value of 93.1\% and 92.7\%, respectively, as mentioned in the beginning of Section \ref{subsec:GCNeval}.

\subsubsection{Hyper-parameter optimization}
\begin{figure}[b]
\vspace{-8mm}
\centering
  \includegraphics[width=0.75\linewidth]{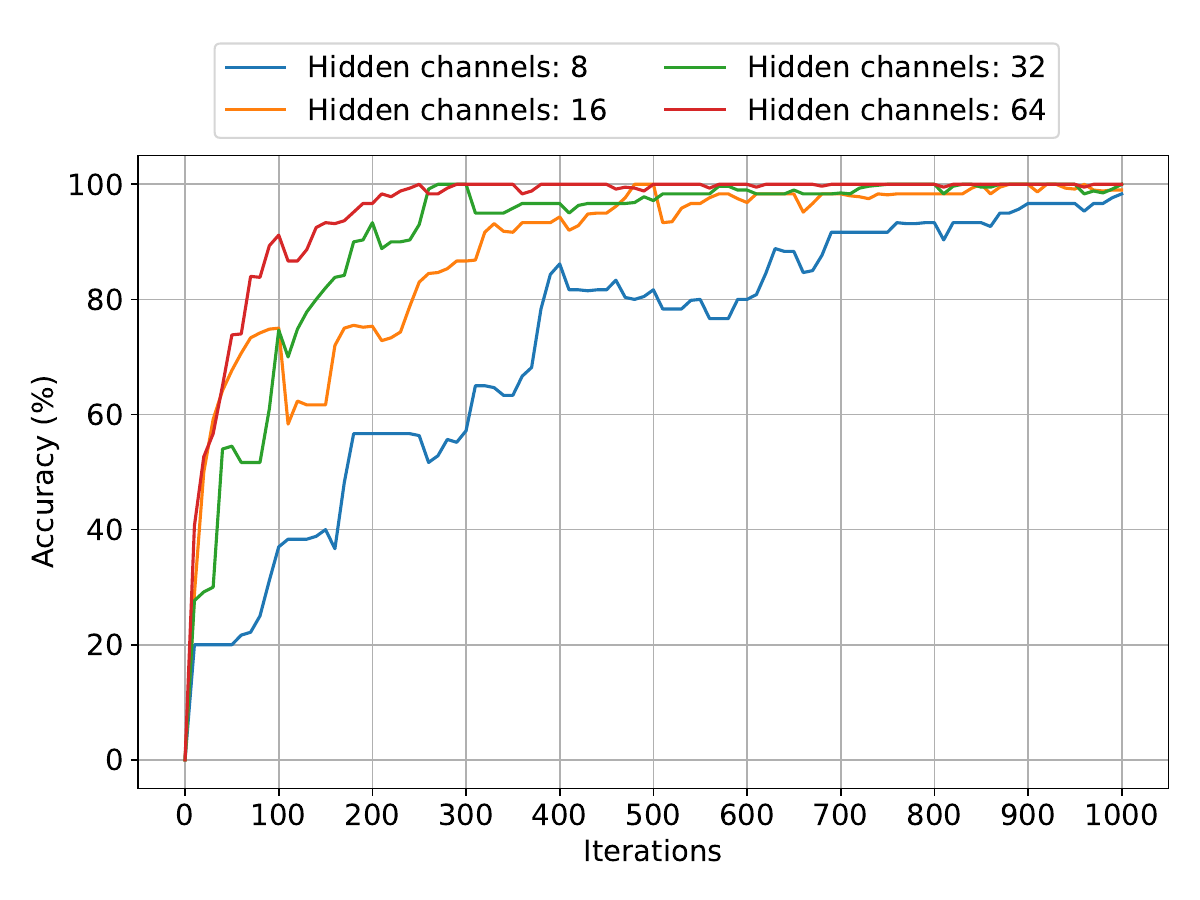}
  \caption{Average circuit detection accuracy of GCN model with varying model size.}
  \label{fig:hiddenchannel}
  \vspace{-4mm}
\end{figure}

To optimize hyper-parameters for the GCN framework, a comprehensive study is conducted, systematically varying the algorithm design parameters. Specifically, we vary the number of hidden channels employed by our GCN model. This parameter is pivotal, as it dictates the dimensions of the GCN model, thereby influencing the classification performance. The evaluation results are demonstrated in Figure \ref{fig:hiddenchannel}. 
In this figure, the y-axis demonstrates the accuracy, while the x-axis depicts the number of iterations over which the performance of our GCN model is evaluated while varying the number of hidden channels in intervals 8, 16, 32, and 64.

From this Figure, it can be observed that the GCN model with 64 hidden channels achieves high accuracy in under 300 iterations, while models with 32, 16, and 8 hidden channels require 500, 600, and over 1000 iterations, respectively, to reach similar accuracy. This indicates that more hidden channels lead to faster convergence due to increased model expressivity. However, we limit the hidden channels to 64, as further increases provide minimal accuracy gains while significantly increasing model complexity.
Therefore, we select the GCN model with 64 hidden channels for our experiments, which furnishes an accuracy of 85.6\% with a precision and recall value of 93.1\% and 92.7\%, respectively (mentioned initially in Section \ref{subsec:GCNeval}).

\subsubsection{Fuzziness} \label{subsec:fuzz}
In our outlined GCN framework, we have operated under the assumption that the detection of CNOT gates between two qubits is flawless. However, evaluations on IBM quantum computers shown in Figure~\ref{fig:guadalupe_data} indicate that perfect detection may not always be achievable, potentially resulting in variations or fuzziness within the data. Consequently, the observed count of detected CNOT gates for specific layers may deviate from the actual count of CNOT gates present, introducing uncertainties into the dataset. In light of this consideration, it becomes imperative to assess the robustness of our GCN model in the presence of such errors.
To address this concern, we have created two additional datasets alongside the one utilized in the preceding analysis, consisting of 336 benchmark circuits. These supplementary datasets consist of 10 and 25 circuits respectively, each constructed using benchmark circuits selected from the broader pool of 336 circuits available. This is performed to facilitate a comprehensive evaluation of the model's robustness against fuzziness across datasets of varying sizes, thereby enriching the validation process.

 

\begin{figure}[t]
\centering
  \includegraphics[width=0.75\linewidth]{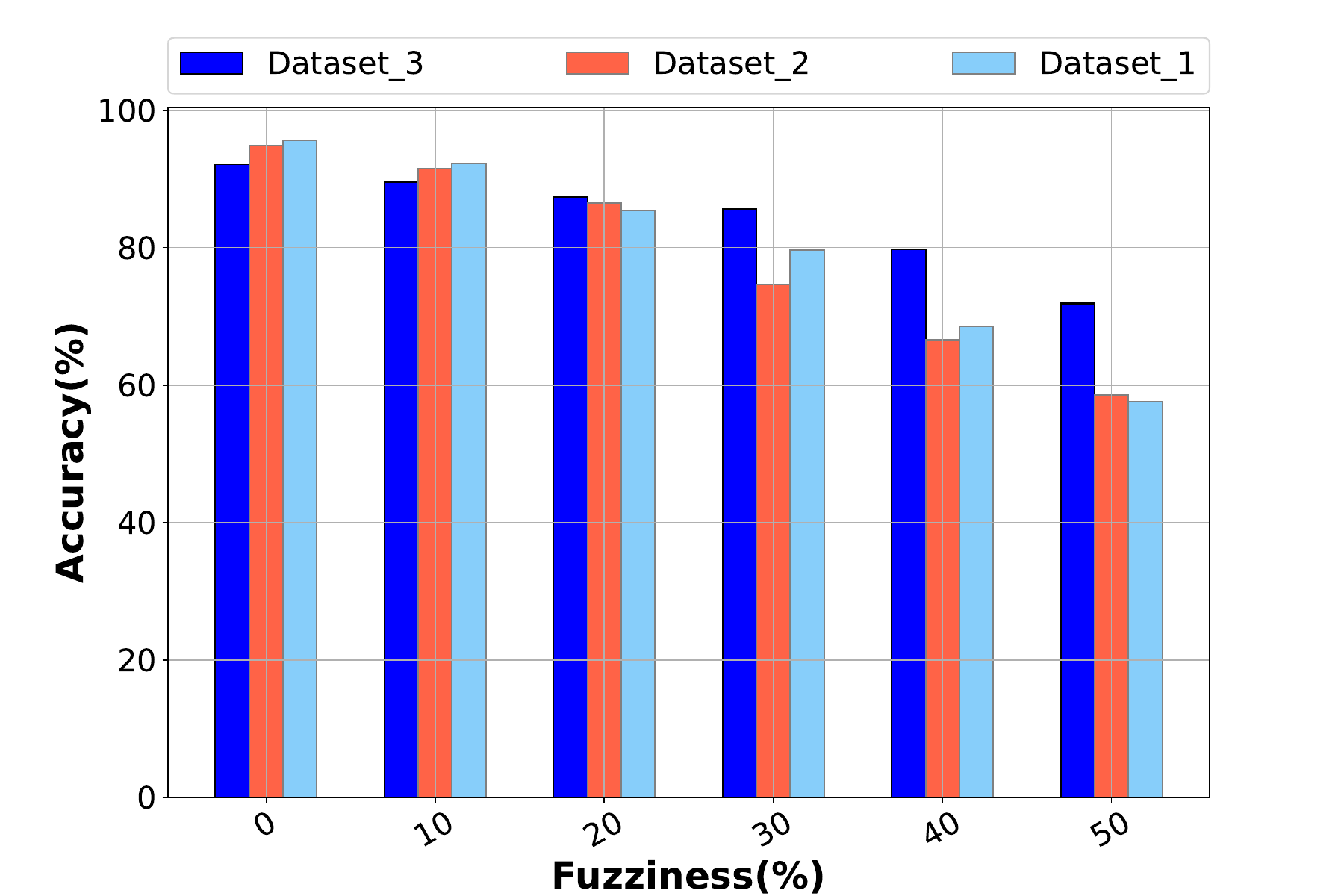}
  \caption{Robustness of proposed GCN model for circuit identification under varying levels of fuzziness in data.}
  \label{fig:fuzziness}
  \vspace{-4mm}
\end{figure}

Figure \ref{fig:fuzziness} provides a visual representation of the robustness exhibited by our GCN model across the three distinct datasets, as influenced by varying levels of fuzziness within the dataset.
In this figure, the x-axis delineates the degree of fuzziness inherent in the dataset, while the y-axis portrays the corresponding accuracy achieved by the model. Specifically,  Dataset\_1 comprises 10 circuits, Dataset\_2 encompasses 25 circuits, and Dataset\_3 corresponds to the dataset derived from 336 benchmark circuits, thereby offering a graduated scale for evaluating the model's performance across datasets of varying complexities and sizes.


To induce fuzziness in our datasets, we introduce a reasonable degree of uncertainty in our CNOT gate detection scheme from an experimental standpoint. Specifically, we allow for a deviation of up to 2 CNOT gates, on each qubit, for each timing layer. This implies that if there is no CNOT gate in a particular timing window, the number of CNOT gates we are allowing to be erroneously detected is two. This fuzziness is randomly included in the dataset to reduce bias.  
This controlled fuzziness spans from 0\% to 50\%,  representing varying levels of error, as shown in Figure \ref{fig:fuzziness}.

From this figure, it is evident that in the presence of fuzziness, the accuracy of our GCN model experiences a consistent decline across all datasets, with a concomitant increase in fuzziness. 
Dataset\_1 and Dataset\_2 achieve accuracies of 93.2\% and 94.5\%, compared to Dataset\_3 which demonstrates a slightly lower accuracy of 90.2\%, for an ideal environment with no fuzziness. However, when exposed to increased levels of fuzziness, Dataset\_3 exhibits greater robustness compared to Dataset\_1 and Dataset\_2. For instance, at an exaggerated fuzziness level of 50\%, the GCN model constructed for Dataset\_3 achieves an accuracy of 70.6\%, outperforming the accuracies yielded by models trained on the other datasets. This heightened robustness can be attributed to the fact that Dataset\_3 is significantly more comprehensive compared to Dataset\_1 and Dataset\_2. When considering a more realistic scenario wherein fuzziness is restricted to 10\%, our model achieves an accuracy of 86.2\% for Dataset\_3, underscoring the robustness of our proposed GCN model and its efficacy in handling real-world scenarios.

\subsection{Performance of different ML Classifiers} \label{subsec:ablation}
To justify our rationale for selecting a GCN-based classifier model, we study the performance of multiple widely employed classification models with the aim of discerning the victim circuit based on information extracted from the adversarial qubits. The outcomes derived from this investigation substantiate the efficacy of our GCN-based classification model.

\paragraph{Dataset}
To evaluate our classifiers, we utilize a dataset as depicted in Table \ref{tab:ev_bms}. These distinct circuits are relatively large, with large gate depth, each comprising 64 qubits. The rationale for selecting these circuits with identical qubit count is to accurately evaluate the performance of classification models while avoiding the imposition of intricate data complexities.
Similar to our complete dataset, this contains 16 transpiled variations for each benchmark circuit (by fixing the compiler optimization level to three and the scheduling method to `as-late-as-possible'). Consequently, with 16 circuits for each benchmark, the dataset comprises 64 circuits.  

\paragraph{Classifiers} The classifiers utilized in our experimental evaluation involve Naive Bayes (\textit{NB}), Support Vector Machine (\textit{SVM}), Random Forest (\textit{RF}), Decision Tree (\textit{DT}), Multi-Layer Perceptron (\textit{MLP}), and Logistic Regression (\textit{LR})~\cite{Suthaharan2016, 6498972,10.1007/978-3-030-03146-6_86, 1687930, HAJMEER200343}. 
These classifiers have been extensively used throughout numerous research works, effectively demonstrating their utility in adeptly classifying comprehensive datasets~\cite{9424362}. For this purpose, following depictions by existing research, we employ these prevalently utilized classifiers to verify their performance in accurately identifying the victim circuit.


\begin{table}[t]
\centering
\caption{ Evaluation benchmarks }
\vspace{-0.05in}
\resizebox{0.9\columnwidth}{!}{%
\begin{tabular}{cccc}
\toprule
      Benchmark           & No. of CNOT Gates    & Gate Depth & No. of Qubits\\
\midrule
qft\_n64          & 259                         & 131                   & 64                         \\ 
adder\_n64             & 175                          & 74                    & 64                         \\ 
ghz\_n64            & 63                          & 65                    & 64                         \\ 
cc\_n64            & 64                          & 195                    & 64                         \\ 
\bottomrule
\end{tabular}%
}
\label{tab:ev_bms}
\vspace{-0.2in}
\end{table}

\paragraph{Results} 

Upon evaluation, our GCN model outperforms other classifiers by over 56\% due to its ability to discern information within qubits and their interconnections, forming a graph-like structure. This enables the GCN model to achieve an impressive 88.2\% accuracy.
\subsection{Discussion of Attack feasibility}


In our experiments, we focus on a quantum computer shared by two users: a victim and an adversary. This two-user model is practical for studying side-channel attacks, as it simplifies the scenario where the adversary can directly target the victim’s circuit. With fewer qubits in use, the adversary can more easily exploit crosstalk to isolate signals from the victim, allowing for clearer detection of side-channel information.

Experimentally demonstrating this attack for more than two users introduces additional challenges. As more users share the quantum computer, increased qubit usage and cross-user interactions intensify crosstalk and noise, making it more difficult for the adversary to isolate the victim’s signals. Moreover, with three or more users, synchronizing sensing circuits with the victim's CNOTs becomes more challenging, requiring probabilistic methods to ensure overlap with the victim’s operations. However, increased crosstalk and attack surface can enable greater opportunities for data leakage for colluding adversaries.


\section{Proposed Defenses} \label{sec:def}

\noindent{\textbf{Advanced Qubit Mapping and Fuzzing:}} Advanced qubit mapping and fuzzing techniques enhance security by obscuring quantum circuit execution, making it harder for adversaries to accurately count CNOT gates. Dynamic qubit reallocation~\cite{meijer2023dynamic} periodically remaps jobs to different qubit sets, adding spatial unpredictability and disrupting adversarial analysis. Randomizing gate sequences further complicates tracking of CNOT execution timings, introducing temporal fuzziness. Instead of reallocating at the job level, cloud providers can diversify qubit usage by executing $N$ circuit shots on different qubits rather than reusing the same qubits, reducing exploitable patterns. Prior research shows that using an ensemble of qubit mappings decreases correlated errors and improves NISQ circuit fidelity~\cite{tannu2019ensemble}. However, randomized mapping complicates resource management, as circuits must be compiled for multiple qubit subsets without adding complexity. Colluding users might still manipulate allocations for co-location, and constantly changing qubit mappings can disrupt error mitigation strategies that depend on consistent noise profiles~\cite{van2023probabilistic}.

\noindent{\textbf{Compiler-assisted Circuit Obfuscation:}} To mask quantum circuits, the compiler can insert dummy gates and padding to confound attempts at reverse-engineering by blurring the distinction between real and false gates. A carefully selected pair of quantum gates can cancel out without affecting the final outcome, making it easy to introduce these dummy operations. Additionally, by strategically adding highly reliable single-qubit gates, CNOT gates can be shifted within the circuit, further complicating any attacker's ability to infer the original structure. However, additional gates can cause more errors, reducing the reliability of user circuits. Instead, we can strategically add single qubit gates. Adding dummy single-qubit gates is straightforward for the compiler, as it doesn't interfere with key transpilation processes like qubit mapping and routing.




\section{Conclusion} \label{sec:Conclusion}

In this paper, we present a novel quantum side-channel attack framework to identify victim circuits in multi-tenant quantum environments by exploiting crosstalk in NISQ systems, exposing vulnerabilities in shared quantum computing.

Our proposed threat model relies on minimal and realistic assumptions validated through experiments that leverage adversarial qubits under specified constraints. These experiments demonstrate successful crosstalk detection from victim circuit qubits. The crosstalk is utilized to extract the spatial and temporal distribution of CNOT gates from this crosstalk.
Finally, we apply a GCN-based model trained to interpret the encoded information within these distributions. When evaluated using 336 benchmark circuits, the proposed attack achieves an accuracy of 85.6\% in identifying the victim circuit.
Our research emphasizes the crucial necessity for strong security measures in expanding collaborative computational environments in future quantum computing paradigms. This paper aims to showcase the threats posed by side channels on quantum computers, laying the groundwork for improving security protocols and strategies to mitigate risks, thereby ensuring the integrity and security of quantum computing technologies.

\section{Ethical considerations}
The experiments described in the paper have been conducted on IBM Quantum computers as discussed in Section \ref{subsec:scademo} and hence, are practically applicable to existing IBM quantum hardware. Our experiments were conducted in controlled environments, carefully replicating quantum computer multi-tenancy, while adhering to strict ethical standards to protect external users' data and confidentiality. By isolating our research, we aimed to advance our understanding of quantum security without compromising ethical standards.

\section{Acknowledgement}
We thank Dr. Adil Ahmad for his insight and contribution to this paper. Kanad Basu, Navnil Choudhury, and Sanjay Das are supported by NSF grant \#2228725. Swamit Tannu and Chaithanya Naik Mude were supported by NSF grants \#2332405 and \#2212232.

\balance

\balance
\bibliographystyle{IEEEtran}
\bibliography{References}

\newpage
\appendix


\section*{ Transpilation of Quantum circuits}

Transpilation is a critical process in quantum computing that converts high-level quantum circuits into hardware-compatible versions, optimized for specific quantum devices. In Qiskit, the transpilation process involves several stages: \textit{layout}, \textit{routing}, \textit{optimization}, and \textit{scheduling}. Each stage helps to ensure the circuit is efficiently executed on real hardware while maintaining the desired fidelity and performance.

\subsection{Layout Stage}
The layout stage determines how logical qubits are mapped to physical qubits on the quantum hardware. Qiskit provides multiple algorithms for the layout stage:

\begin{itemize}
    \item \textbf{TrivialLayout}: Maps logical qubits directly to physical qubits with no optimization.
    \item \textbf{DenseLayout}: Assigns logical qubits to a tightly connected subset of physical qubits, maximizing proximity.
    \item \textbf{NoiseAdaptiveLayout}: Maps logical qubits based on hardware error rates, placing critical qubits on those with the lowest noise.
    \item \textbf{SabreLayout}: Dynamically optimizes the layout to minimize SWAP operations as the circuit evolves.
\end{itemize}
A good layout reduces the need for qubit swaps, lowering circuit depth and potential errors.

\subsection{Routing Stage}
The routing stage ensures that qubits can interact even when not physically adjacent by inserting SWAP gates. Qiskit offers various routing algorithms:
\begin{itemize}
    \item \textbf{BasicSwap}: A straightforward approach that adds SWAP gates wherever needed to enable qubit interaction.
    \item \textbf{LookaheadSwap}: Looks ahead in the circuit to minimize the number of SWAP gates.
    \item \textbf{StochasticSwap}: Uses probabilistic methods to find more efficient SWAP gate placement.
    \item \textbf{SabreSwap}: Optimizes SWAP gate insertion dynamically, minimizing both gate count and depth.
\end{itemize}
Routing is important for managing qubit connectivity constraints on physical quantum hardware while keeping the circuit as compact as possible.

\subsection{Optimization Level}
Qiskit allows for four levels of optimization during transpilation, balancing circuit fidelity and efficiency:
\begin{itemize}
    \item \textbf{Level 0}: No optimization; the circuit is directly mapped to the hardware.
    \item \textbf{Level 1}: Light optimization, focusing on basic reductions like gate cancellations.
    \item \textbf{Level 2}: More advanced optimizations that reduce gate count and circuit depth.
    \item \textbf{Level 3}: Aggressive optimizations, such as gate re-synthesis and fusion, often result in the most efficient circuit representation.
\end{itemize}
While higher optimization levels improve the performance of quantum circuits on hardware, they require more computational resources during the transpilation process.

\subsection{Scheduling Stage}
\begin{figure}[b]
\centering
  \includegraphics[width=\linewidth]{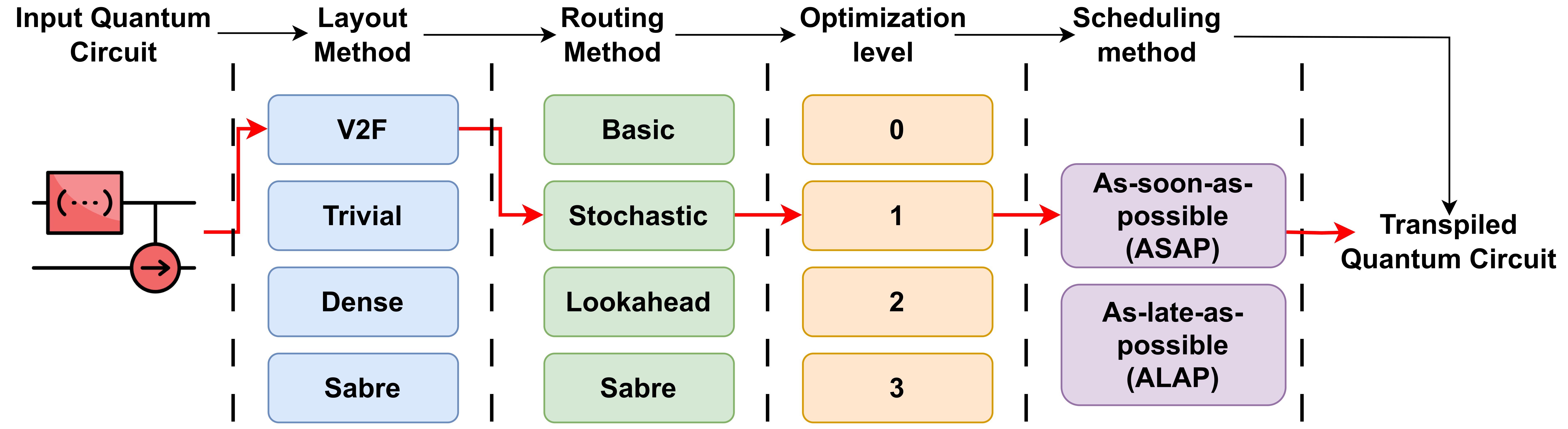}
  \caption{Different possible transpilation options for a single quantum circuit.}
  \label{fig:transpilation}
  \vspace{-8mm}
\end{figure}

The scheduling stage determines when each quantum gate should be executed, considering the hardware's timing constraints. Qiskit supports two scheduling methods:
\begin{itemize}
    \item \textbf{ASAP (As Soon As Possible)}: Executes gates as early as possible to minimize circuit runtime.
    \item \textbf{ALAP (As Late As Possible)}: Delays gate execution to minimize idle qubit time, reducing decoherence effects.
\end{itemize}
Scheduling is essential for optimizing the total execution time and minimizing errors due to qubit decoherence.
The various options of transpilation for a single quantum circuit are shown in Figure \ref{fig:transpilation}, where it can be seen that, based on the options selected for each transpilation stage, a single quantum circuit can be transpiled in 128 distinct ways.

Furthermore, to elucidate the difference in the structure of a quantum circuit based on the transpilation configurations is shown in Table \ref{tab:trans}. In the table, the first column represents the quantum circuit undergoing transpilation. The second and third columns represent the layout and routing stages selected, respectively. Columns four to eight represent the number of CNOT gates present in each time bucket, where each time bucket has a uniform duration. Finally, column nine depicts the total number of CNOT gates for each configuration.

\begin{table*}[t!]
\centering
\caption{Difference in Transpilation Configurations}
\label{tab:trans}
\resizebox{0.9\textwidth}{!}{%
\begin{tabular}{|c|cc|cccccc|}
\hline
\multirow{2}{*}{\textbf{Quantum circuit}} &
  \multicolumn{2}{c|}{\textbf{Transpilation configuration}} &
  \multicolumn{6}{c|}{\textbf{CNOT count}} \\ \cline{2-9} 
 &
  \multicolumn{1}{c|}{\textbf{Layout method}} &
  \textbf{Routing method} &
  \multicolumn{1}{c|}{\textbf{Time bucket 1}} &
  \multicolumn{1}{c|}{\textbf{Time bucket 2}} &
  \multicolumn{1}{c|}{\textbf{Time bucket 3}} &
  \multicolumn{1}{c|}{\textbf{Time bucket 4}} &
  \multicolumn{1}{c|}{\textbf{Time bucket 5}} &
  \textbf{Total CNOT gates} \\ \hline
\multirow{16}{*}{\textbf{QAOA}} &
  \multicolumn{1}{c|}{\multirow{4}{*}{Trivial}} &
  Basic &
  \multicolumn{1}{c|}{12} &
  \multicolumn{1}{c|}{14} &
  \multicolumn{1}{c|}{10} &
  \multicolumn{1}{c|}{15} &
  \multicolumn{1}{c|}{10} &
  61 \\ \cline{3-9} 
 &
  \multicolumn{1}{c|}{} &
  Stochastic &
  \multicolumn{1}{c|}{11} &
  \multicolumn{1}{c|}{12} &
  \multicolumn{1}{c|}{9} &
  \multicolumn{1}{c|}{14} &
  \multicolumn{1}{c|}{8} &
  54 \\ \cline{3-9} 
 &
  \multicolumn{1}{c|}{} &
  Lookahead &
  \multicolumn{1}{c|}{13} &
  \multicolumn{1}{c|}{12} &
  \multicolumn{1}{c|}{11} &
  \multicolumn{1}{c|}{16} &
  \multicolumn{1}{c|}{9} &
  61 \\ \cline{3-9} 
 &
  \multicolumn{1}{c|}{} &
  Sabre &
  \multicolumn{1}{c|}{10} &
  \multicolumn{1}{c|}{13} &
  \multicolumn{1}{c|}{10} &
  \multicolumn{1}{c|}{15} &
  \multicolumn{1}{c|}{9} &
  57 \\ \cline{2-9} 
 &
  \multicolumn{1}{c|}{\multirow{4}{*}{Dense}} &
  Basic &
  \multicolumn{1}{c|}{15} &
  \multicolumn{1}{c|}{16} &
  \multicolumn{1}{c|}{14} &
  \multicolumn{1}{c|}{13} &
  \multicolumn{1}{c|}{11} &
  69 \\ \cline{3-9} 
 &
  \multicolumn{1}{c|}{} &
  Stochastic &
  \multicolumn{1}{c|}{14} &
  \multicolumn{1}{c|}{15} &
  \multicolumn{1}{c|}{13} &
  \multicolumn{1}{c|}{14} &
  \multicolumn{1}{c|}{10} &
  66 \\ \cline{3-9} 
 &
  \multicolumn{1}{c|}{} &
  Lookahead &
  \multicolumn{1}{c|}{14} &
  \multicolumn{1}{c|}{17} &
  \multicolumn{1}{c|}{14} &
  \multicolumn{1}{c|}{12} &
  \multicolumn{1}{c|}{12} &
  69 \\ \cline{3-9} 
 &
  \multicolumn{1}{c|}{} &
  Sabre &
  \multicolumn{1}{c|}{13} &
  \multicolumn{1}{c|}{14} &
  \multicolumn{1}{c|}{12} &
  \multicolumn{1}{c|}{13} &
  \multicolumn{1}{c|}{11} &
  63 \\ \cline{2-9} 
 &
  \multicolumn{1}{c|}{\multirow{4}{*}{V2F}} &
  Basic &
  \multicolumn{1}{c|}{10} &
  \multicolumn{1}{c|}{13} &
  \multicolumn{1}{c|}{11} &
  \multicolumn{1}{c|}{12} &
  \multicolumn{1}{c|}{9} &
  55 \\ \cline{3-9} 
 &
  \multicolumn{1}{c|}{} &
  Stochastic &
  \multicolumn{1}{c|}{9} &
  \multicolumn{1}{c|}{12} &
  \multicolumn{1}{c|}{10} &
  \multicolumn{1}{c|}{12} &
  \multicolumn{1}{c|}{8} &
  51 \\ \cline{3-9} 
 &
  \multicolumn{1}{c|}{} &
  Lookahead &
  \multicolumn{1}{c|}{11} &
  \multicolumn{1}{c|}{13} &
  \multicolumn{1}{c|}{10} &
  \multicolumn{1}{c|}{13} &
  \multicolumn{1}{c|}{9} &
  56 \\ \cline{3-9} 
 &
  \multicolumn{1}{c|}{} &
  Sabre &
  \multicolumn{1}{c|}{9} &
  \multicolumn{1}{c|}{11} &
  \multicolumn{1}{c|}{9} &
  \multicolumn{1}{c|}{12} &
  \multicolumn{1}{c|}{8} &
  49 \\ \cline{2-9} 
 &
  \multicolumn{1}{c|}{\multirow{4}{*}{Sabre}} &
  Basic &
  \multicolumn{1}{c|}{13} &
  \multicolumn{1}{c|}{14} &
  \multicolumn{1}{c|}{12} &
  \multicolumn{1}{c|}{15} &
  \multicolumn{1}{c|}{10} &
  64 \\ \cline{3-9} 
 &
  \multicolumn{1}{c|}{} &
  Stochastic &
  \multicolumn{1}{c|}{12} &
  \multicolumn{1}{c|}{13} &
  \multicolumn{1}{c|}{11} &
  \multicolumn{1}{c|}{13} &
  \multicolumn{1}{c|}{9} &
  58 \\ \cline{3-9} 
 &
  \multicolumn{1}{c|}{} &
  Lookahead &
  \multicolumn{1}{c|}{14} &
  \multicolumn{1}{c|}{15} &
  \multicolumn{1}{c|}{12} &
  \multicolumn{1}{c|}{14} &
  \multicolumn{1}{c|}{11} &
  66 \\ \cline{3-9} 
 &
  \multicolumn{1}{c|}{} &
  Sabre &
  \multicolumn{1}{c|}{12} &
  \multicolumn{1}{c|}{13} &
  \multicolumn{1}{c|}{11} &
  \multicolumn{1}{c|}{14} &
  \multicolumn{1}{c|}{9} &
  59 \\ \hline
\end{tabular}%
}
\end{table*}

From the table, it becomes evident that starting with the same quantum algorithm, represented by the initial quantum circuit, the choice of transpilation configurations—especially the layout and routing strategies—can significantly influence the final structure and performance of the circuit. These changes manifest not only in the gate count but also in the spatial and temporal arrangement of operations, with each transpilation configuration producing a distinct version of the circuit.

One of the most critical observations from the table is the impact of layout and routing choices on the number of CNOT gates. CNOT gates, being two-qubit operations, are generally more prone to noise and errors compared to single-qubit gates, making them a bottleneck in the execution of quantum algorithms. A quantum circuit with an excessive number of CNOT gates or an inefficient distribution of those gates across time is more likely to suffer from errors, reduced fidelity, and longer execution times. Therefore, minimizing the number of CNOT gates—or at least distributing them efficiently—becomes essential for improving the performance of the circuit.

The layout algorithm selected in the second column of Table \ref{tab:trans} plays a crucial role in determining the initial mapping of logical qubits to physical qubits on the quantum device. Poor qubit mapping may force the transpiler to introduce unnecessary SWAP operations during the routing stage to accommodate the physical constraints of the hardware, increasing the number of two-qubit gates like CNOTs. A well-chosen layout algorithm, such as NoiseAdaptiveLayout, which assigns critical qubits to the least noisy physical qubits, may reduce the need for such SWAP operations and directly improve the circuit's overall fidelity. In contrast, a simple TrivialLayout method, while faster to compute, might not account for hardware-specific noise and connectivity, leading to more cumbersome routing on the quantum hardware and a higher gate count.

In the routing stage (third column), the choice of routing algorithm determines how efficiently non-adjacent qubits are brought together to perform two-qubit gates. For example, a routing algorithm like SabreSwap, which dynamically selects SWAP operations to minimize overall gate costs, can help to significantly reduce the number of additional gates introduced into the circuit. The reduction of SWAP gates is not merely a matter of minimizing gate count; it also helps in reducing circuit depth, which is directly tied to the execution time and the qubits' exposure to decoherence. Conversely, simpler routing algorithms like BasicSwap might lead to an increase in both the number of CNOT gates and the overall circuit depth, as SWAP gates are introduced indiscriminately without regard for long-term circuit efficiency.

The distribution of CNOT gates across time buckets (represented in columns four through eight) provides further insight into the temporal efficiency of the transpiled circuits. Uneven distribution, where a large number of CNOT gates cluster in one or two time buckets, may lead to qubits idling for extended periods, increasing their vulnerability to decoherence and noise. In contrast, a more uniform distribution of CNOT gates across time buckets suggests that the transpiler has balanced the gate execution schedule effectively, thereby minimizing idle times and the cumulative effect of errors. This is especially important for NISQ (Noisy Intermediate-Scale Quantum) devices, where hardware noise and gate errors are major concerns.

The total CNOT count (ninth column) ultimately reflects the overall gate complexity of each transpilation configuration. A higher total number of CNOT gates indicates a more complex circuit, which can negatively impact execution fidelity, particularly on noisy hardware. Circuits with fewer CNOT gates tend to have higher success rates because they are less susceptible to noise accumulation. However, this metric must be considered in conjunction with circuit depth and the distribution of operations. A circuit with fewer CNOT gates but a highly uneven time distribution might still perform worse than one with a slightly higher CNOT count but a more efficient gate scheduling and lower depth. This underscores the importance of striking a balance between the total gate count, the circuit depth, and the temporal distribution of operations.

Ultimately, transpilation configurations produce circuits with diverse structural properties. The ability to control layout, routing, optimization, and scheduling allows users to tailor quantum circuits to specific hardware, balancing the trade-offs between gate count, circuit depth, and execution time. By understanding the impact of these configurations, users can select the best options for their quantum algorithm, given the hardware constraints and the specific goals of their computation. For example, applications requiring higher fidelity may prioritize configurations that minimize CNOT gates and depth, while other tasks may allow for deeper circuits with more operations if execution speed or computational resource constraints are more pressing.

This flexibility is especially important in the current era of quantum computing, where hardware limitations such as qubit connectivity, gate fidelity, and coherence times vary significantly between devices. Therefore, the choice of transpilation configurations is not a one-size-fits-all solution but rather a key decision that can dramatically alter the performance and structure of a quantum circuit based on the user’s objectives and the hardware being used. By carefully analyzing the impact of these configurations, users can ensure that their circuits are optimized for their specific needs, improving the likelihood of successful quantum computation.

\addcontentsline{toc}{section}{Appendix} 
\end{document}